\documentclass[%
 reprint,
superscriptaddress,
nofootinbib,
 amsmath,amssymb,
 aps,
floatfix,
]{revtex4-1}

\usepackage{graphicx}
\usepackage{dcolumn}
\usepackage{bm}
\usepackage[colorlinks=true]{hyperref}
\usepackage[normalem]{ulem}

\begin{document}

\preprint{APS/123-QED}

\title{Periastron precession effect of $f$-mode dynamical tides on gravitational waves from eccentric double white dwarfs}

\author{Shu Yan Lau} \email{sl8ny@virginia.edu}
\affiliation{Department of Physics, University of Virginia, Charlottesville, Virginia 22904, USA}

\author{Kent Yagi} \email{ky5t@virginia.edu}
\affiliation{Department of Physics, University of Virginia, Charlottesville, Virginia 22904, USA}

\author{Phil Arras} \email{pla7y@virginia.edu}
\affiliation{Department of Astronomy, University of Virginia, Charlottesville, Virginia 22904, USA}

\date{\today}

\begin{abstract}
The dynamical tide can play an important role in the orbital motion of close eccentric double white dwarf binaries.
As the launching of the space-based gravitational-wave detector, the Laser Interferometer Space Antenna (LISA), is just around the corner, detection of gravitational wave signals from such systems is anticipated.
In this paper, we discuss the influence of the dynamical tide on eccentric orbits, focusing on the effect on periastron precession. 
We show that in orbits with a high eccentricity, resonance can cause a large precession when a harmonic of the orbital frequency matches the natural frequencies of the normal modes of the star. In contrast to the case with circular orbits, each mode can encounter multiple resonances with different harmonics and these resonant regions can cover about 10\% of the frequency space for orbits with close separations. 
In this case, the tidal precession effect is distinct from the other contributions and can be identified with LISA if the signal-to-noise ratio is high enough. However, within the highly eccentric-small separation region, the dynamical tide causes chaotic motion and the gravitational wave signal becomes unpredictable.
Even not at resonance, the dynamical tide can contribute up to $20$\% of the precession for orbits close to Roche-lobe filling separation with low eccentricities and LISA can resolve these off-resonant dynamical tide effects within the low eccentricity-small orbital separation region of the parameter space. For lower mass systems, the dynamical tide effect can degenerate with the uncertainties of the eccentricity, making it unmeasurable from the precession rate alone. For higher mass systems, the radiation reaction effect becomes significant enough to constrain the eccentricity, allowing the measurement of the dynamical tide. 

\end{abstract}

\maketitle

\section{\label{Sec:intro}Introduction}

The Laser Interferometer Space Antenna (LISA), \cite{2017arXiv170200786A} planned to launch in 2034, covers the  0.1-100~mHz part of the gravitational wave (GW) spectrum, which is expected to contain a variety of sources different from those covered by the current ground-based detectors. Double white dwarf (DWD) systems within the Milky Way Galaxy are expected to be one of the most promising sources among those, and we can expect about $10^4$ of such systems within the resolvable frequency range of LISA \cite{2001A&A...375..890N,10.1093/mnras/stz2834}. The majority of these systems are expected to have circularized during the early evolution stage due to tidal interactions, and the frequency shift from orbital decay within the observation time is also expected to be much smaller than the signal frequency \cite{Takahashi_2002}. Hence, the signal is considered quasi-monochromatic and is mainly characterized by three parameters: the amplitude, frequency, and time derivative of the frequency. For the cases with non-negligible eccentricities, the signal contains multiple harmonics that have additional dependence on the precession rate and the eccentricity \cite{10.1093/mnras/274.1.115}, with the former one modulating the signal amplitude and the latter governing the relative amplitudes of the harmonics.

How do highly eccentric DWDs form?
The major formation channels of isolated DWD systems involve their progenitors going through two mass transfer processes, with one or more common envelope stages \cite{Woods_2011,2000A&A...360.1011N,1993PASP..105.1373I}. The binary has to be close enough and the tidal interaction is expected to circularize the system before forming the two WDs. To form an eccentric DWD, dynamical processes are required. The study by Willems~\textit{et~al.} \cite{Willems_2007} has estimated the population of these eccentric DWD systems formed inside the globular clusters within the Milky Way, based on the simulations of different globular cluster models in \cite{10.1111/j.1365-2966.2006.10876.x}. They conclude that there are at least a handful of such systems within the LISA band that can be detected with more than one harmonic at a signal-to-noise ratio (SNR) higher than 8. Other than that, a close binary of DWD within a hierarchical triple system can also possess a high eccentricity \cite{Thompson_2011, 2013arXiv1305.2191P} through the Kozai-Lidov mechanism \cite{1962AJ.....67..591K, LIDOV1962719}. A rough estimation of the population of these sources is given by Gould \cite{Gould_2011}, stating that there may be $\sim200$ of these sources within 1~kpc detectable by LISA. However, more detailed triple system parameters and orbital evolution models are required to obtain an accurate estimate of the population.

The argument of the pericenter of an eccentric orbit precesses due to the non-Keplerian effects such as the corrections from general relativity (GR). This results in a slow amplitude modulation on the waveform amplitude, splitting each harmonic into three modes. The precession rate can be measured from the signal as half of the frequency difference of the splitting. This has been demonstrated in Willems~\textit{et~al.} \cite{PhysRevLett.100.041102}, which studies the precession from the tide and the so-called first post-Newtonian order (1PN) leading GR effect. In particular, the tidal effect is treated in the static limit and has shown a significant contribution to the precession of the orbit at intermediate separations. The full tidal contribution, however, contains also a dynamical part that depends on the excitation of individual oscillation modes of the WDs and can also affect the orbit in various ways.

Previous studies have shown the dynamical tide effects on the orbital motion in different time scales.
In McNeill \textit{et al.} \cite{10.1093/mnras/stz3215}, the dynamical tide has been shown to induce eccentricities on circular DWD orbits and give rise to non-zero amplitudes in the first and third harmonics of the orbital frequency in the emitted GW signal. On the orbital evolution, Fuller and Lai \cite{10.1111/j.1365-2966.2010.18017.x, 10.1111/j.1365-2966.2011.20320.x}, and Burkart \textit{et al.} \cite{10.1093/mnras/stt726} considered the effect of the dissipation from gravity modes ($g$-modes) on the tidal synchronization. The latter has examined the resonance locking \cite{1999A&A...350..129W} in which the system stays near the resonance of one oscillation mode. Dynamical tide has also been shown to cause chaotic motion in highly eccentric systems with small pericenter separations \cite{1995ApJ...450..732M}.

In this study, we focus on the effect of the dynamical tide on the orbital precession. We derive the formula of the precession rate caused by the full tidal effect using the method of osculating orbit. This allows us to compare it with the other contributing factors to the precession rate. We then explore the parameter space where such an effect becomes significant and estimate the influence on the signal detected by LISA. 

The paper is organized as follows: In Sec.~\ref{Sec:dyn_tide_formulation}, we briefly introduce the formulation of the linear tidal problem. Then, we discuss the results of the tidal precession on the orbit in Sec.~\ref{Sec:Precession}. In Sec.~\ref{Sec:Precession_on_waveform}, we discuss the effect of precession caused by the dynamical tide on waveform detection by LISA. Lastly, we summarize our findings in Sec.~\ref{Sec:conclusion} and provide potential future avenues.

We use the following notation throughout this paper:
We define $m_1$, $R_1$ to be the mass and radius of one of the WDs (denoted as WD 1), and $m_2$, $R_2$ for those of the other one (WD 2).

\section{\label{Sec:dyn_tide_formulation} The formulation of the tidal problem}

The classical Lagrangian perturbation theory \cite{1967MNRAS.136..293L,1978ApJ...221..937F} allows us to describe the fluid motion inside the WD at any instant in terms of the Lagrangian displacement vector $\boldsymbol{\xi}(t,\mathbf{x})$. In the center-of-mass frame of the WD 1, the position vector of a fluid element in the perturbed star is given by $\mathbf{x} + \boldsymbol{\xi}(t,\mathbf{x})$, where $\mathbf{x}$ is the original position of the fluid element in the unperturbed WD. The vector $\boldsymbol{\xi}(t,\mathbf{x})$ satisfies the equation of motion:
\begin{align}
    \rho \ddot{\boldsymbol{\xi}} = \boldsymbol{f}[\boldsymbol{\xi}] -\rho \boldsymbol{\nabla} U, \label{eq:xi_eom}
\end{align}
where $\boldsymbol{f}[\boldsymbol{\xi}]$ represents the internal restoring force with respect to the equilibrium configuration, $\rho$ is the density and $U$ is the tidal potential due to WD 2. We follow the formulation in \cite{PhysRevD.65.024001} to determine the tidal response and the corresponding back-reaction on the orbit. In this section, we briefly review the main equations. For further details, we refer the readers to \cite{PhysRevD.65.024001,Weinberg_2012}.

We first focus on the deformation of WD 1. The induced quadrupolar deformation causes WD 1 to exert an extra force in addition to the point-mass contribution onto the orbit. The contributions from WD 2 on the orbital motion are completely symmetric to that from WD 1 and can be found by switching the labels, as well as noting that the azimuthal angle $\Phi$ from WD 2 differs by a phase of $\pi$.

Following \cite{PhysRevD.65.024001}, we expand the phase space vector as 
\begin{align}
    \begin{bmatrix}
        \boldsymbol{\xi} \\
        \dot{\boldsymbol{\xi}}
    \end{bmatrix}
    = \sum_\alpha q_\alpha(t)
    \begin{bmatrix}
        \boldsymbol{\xi}_\alpha(\mathbf{x}) \\
        - i \omega_\alpha \boldsymbol{\xi}_\alpha(\mathbf{x})
    \end{bmatrix}. \label{eq:phase_space_eigenmode}
\end{align}
Here, $\boldsymbol{\xi}_\alpha$ represents an eigenmode with an eigenfrequency $\omega_\alpha$, $q_\alpha(t)$ is the excitation amplitude of the mode and the subscript $\alpha$ represents the set of quantum numbers that specifies an eigenmode, as well as the sign of the frequency to account for a phase space mode and its complex conjugate. 
The eigenvalue problem is written as 
\begin{align}
    -\rho \omega_\alpha^2 \boldsymbol{\xi}_\alpha = \boldsymbol{f}[\boldsymbol{\xi}_\alpha].
\label{eq:xi_eom_eigen}
\end{align}
The eigenmodes are normalized such that \begin{align}
    2 \omega_\alpha^2 \int d^3x \rho \boldsymbol{\xi}_\alpha^* \cdot \boldsymbol{\xi}_\alpha = \frac{G m_1^2}{R_1}.
\end{align}
The detail of Eq.~\eqref{eq:xi_eom_eigen} and the eigenvalue problem can be found in e.g., \cite{1971AcA....21..289D, 1989nos..book.....U}.
In Table~\ref{Tab:WD_models}, we list the parameters of the WDs in this study, which are constructed using the zero temperature equation of state of a degenerate electron gas and ions with mean molecular weight per electron $\mu_e = 2$. The important non-radial mode, known as the fundamental mode ($f$-mode), dominates the tidal deformation at the quadrupolar order ($l=2$) in such models. For simplicity, we consider only the contributions from the $l=2$ $f$-mode in the following as the overlap integrals of other modes, e.g., the pressure modes ($p$-modes), are smaller by at least one order of magnitude.

\begin{table}
\begin{ruledtabular}
\begin{tabular}{cccc}
mass & radius & $\omega_\alpha$ & $I_{\alpha l m}$ \\ 
$M_\odot$ & $10^3$~km & s$^{-1}$ &\\\hline
0.20 & 14.66 & 0.135 & 0.234 \\
0.60 & 8.84 & 0.520 & 0.216 
\end{tabular}
\end{ruledtabular}
\caption{\label{Tab:WD_models} WD parameters obtained by solving the eigenvalue problem Eq.~\eqref{eq:xi_eom_eigen}. The eigenfrequency and overlap integral, $\omega_\alpha$ and $I_{\alpha l m}$, of the $l=2$ $f$-modes are listed.
}
\end{table}

In the comoving frame of WD 1, the mode amplitude $q_\alpha(t)$ satisfies the equation \cite{PhysRevD.65.024001, Weinberg_2012}
\begin{align}
    \dot{q}_\alpha +i\omega_\alpha q_\alpha = i\omega_\alpha U_\alpha, \label{eq:mode_amp_eom}
\end{align}
where
\begin{align}
    U_\alpha = \frac{m_2}{m_1}\sum_{lm} W_{lm}I_{\alpha l m}\left(\frac{R_1}{D}\right)^{l+1} e^{-im\Phi}. \label{eq:U_alpha}
\end{align}
Here, $D$ is the orbital separation, $\Phi$ is the azimuthal coordinate of WD 2 as seen by WD 1, and $W_{lm}$ is given by $[4\pi/(2l+1)] Y_{lm}\left(\pi/2,0\right)$. We have also defined the overlap integral $I_{\alpha l m}$
\begin{align}
    I_{\alpha l m} = \frac{1}{m_1 R_1^l}\int d^3 x \rho \boldsymbol{\xi}^*_\alpha \cdot \boldsymbol{\nabla}\left[r^l Y_{lm}(\theta,\phi) \right],
\end{align}
where $Y_{lm}$ are the spherical harmonics and $(r,\theta,\phi)$ are the spherical coordinates with the origin at the stellar center of the WD 1. Note that since we have not included the Coriolis force in Eq.~\eqref{eq:xi_eom}, $I_{\alpha l m} = I_{\alpha} \delta_{l, l_\alpha} \delta_{m, m_\alpha}$, where $\delta_{i,j}$ is the Kronecker delta function and $l_\alpha, m_\alpha$ are the spherical harmonic indices contained in $\alpha$.
Following Appendix C of \cite{Weinberg_2012}, the orbital acceleration due to the tidally deformed WDs is given by
\begin{align}
    \mathbf{a}_\text{tide} =& -\frac{GM}{R_1^2} \sum_{\alpha l m}  W_{lm} I_{\alpha lm} \left(\frac{R_1}{D}\right)^{l+2}e^{-i m \Phi} \nonumber\\
    &\times q_\alpha^* \left[(l+1)\mathbf{n}+i m \boldsymbol{\lambda}\right], \label{eq:tide_acc}
\end{align}
where $M = m_1+m_2$, $\mathbf{n}$ is the unit vector in the radial direction from WD 2 to WD 1 and $\boldsymbol{\lambda}$ is the unit vector in the tangential direction.

\section{\label{Sec:Precession} Precession rate due to the dynamical tide}

The dynamical tide provides a non-Keplerian force that leads to the precession of the pericenter. Willems \textit{et al.} \cite{PhysRevLett.100.041102} have studied precession in DWD systems due to equilibrium tide, rotation, and 1PN correction and have shown that detecting the various precession effects with LISA allows one to probe the interior structure of WDs (also see \cite{Valsecchi_2012} which extends the work by considering more detailed WD models). Unlike these factors that have a relatively simple dependence on the parameters of the WD or the orbit, the effect of dynamical tide depends on the details of the interplay between the oscillation modes and the orbital motion. In this section, we investigate the precession rate due to the dynamical tide and its effects on the GW signal.

\begin{figure*}
    \includegraphics[width=8.6cm]{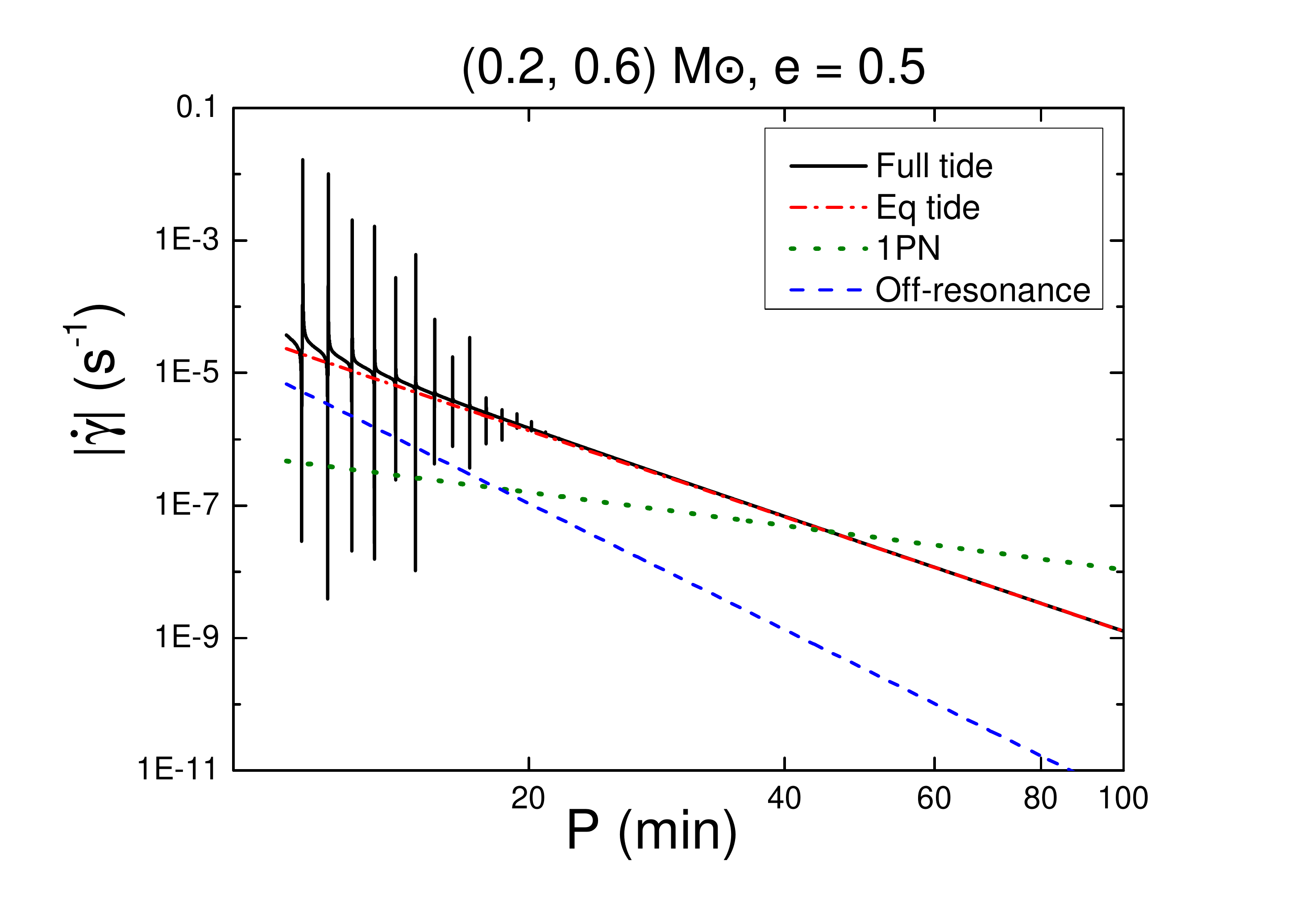}
    \includegraphics[width=8.6cm]{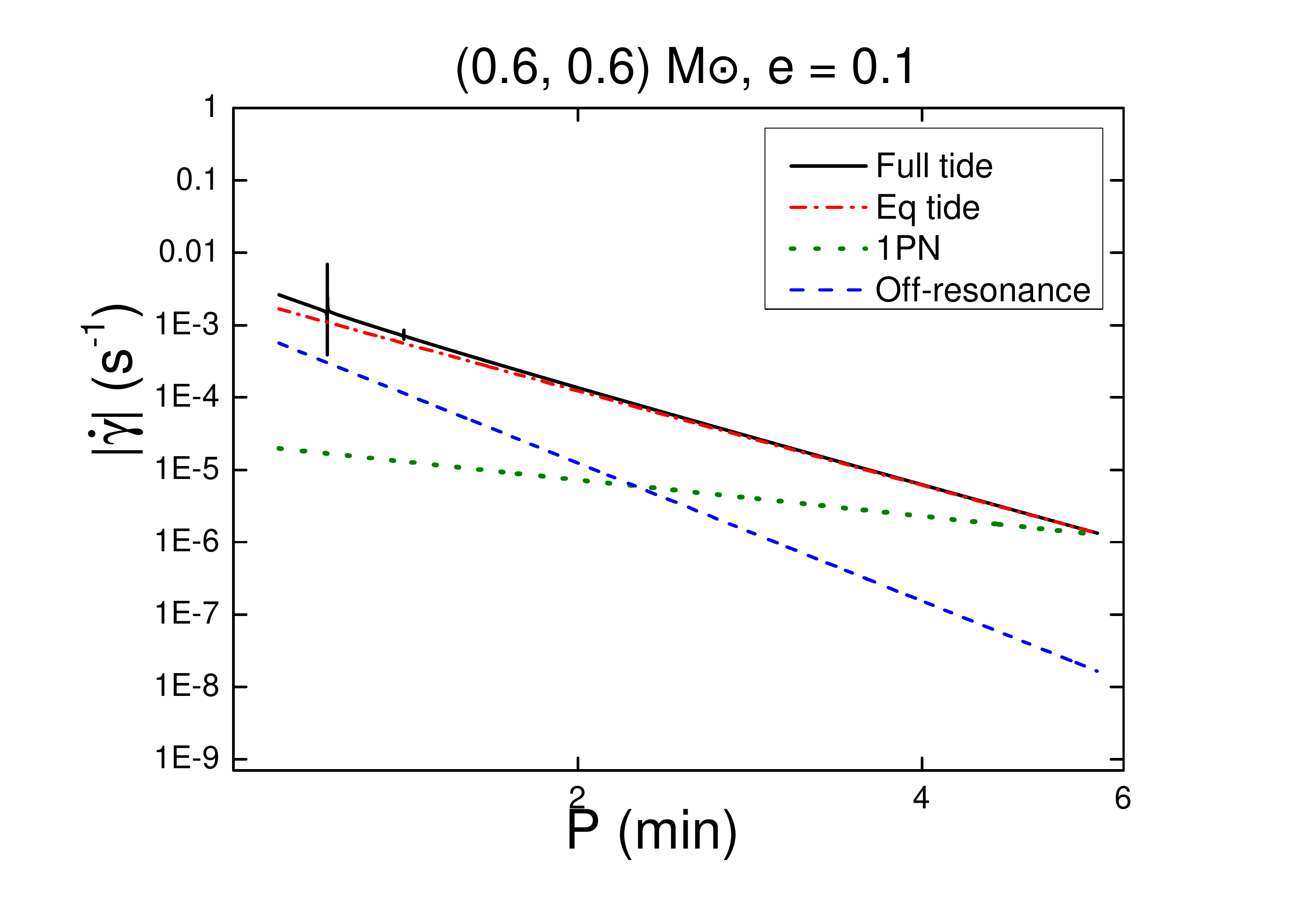}
    \includegraphics[width=8.6cm]{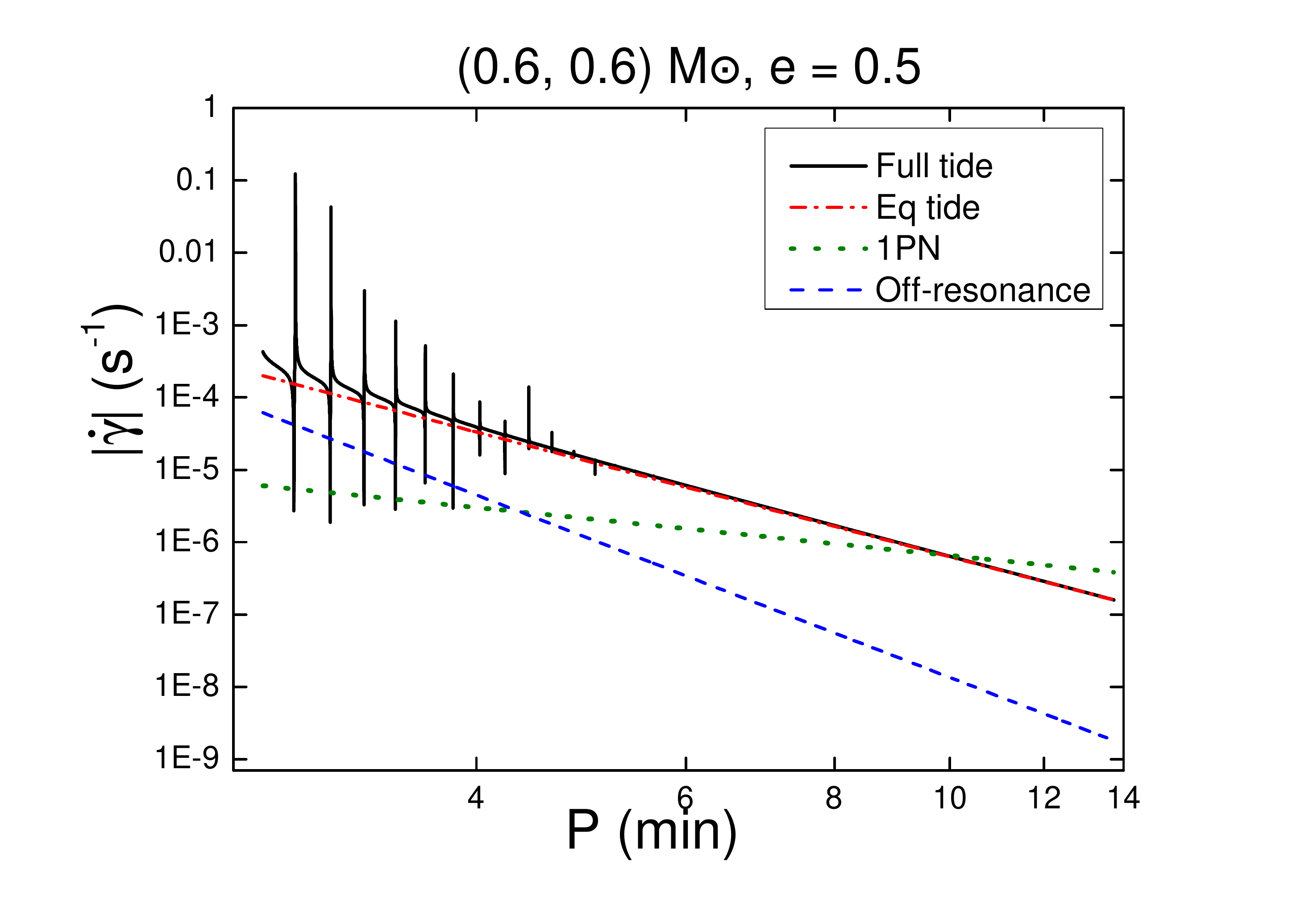}
    \includegraphics[width=8.6cm]{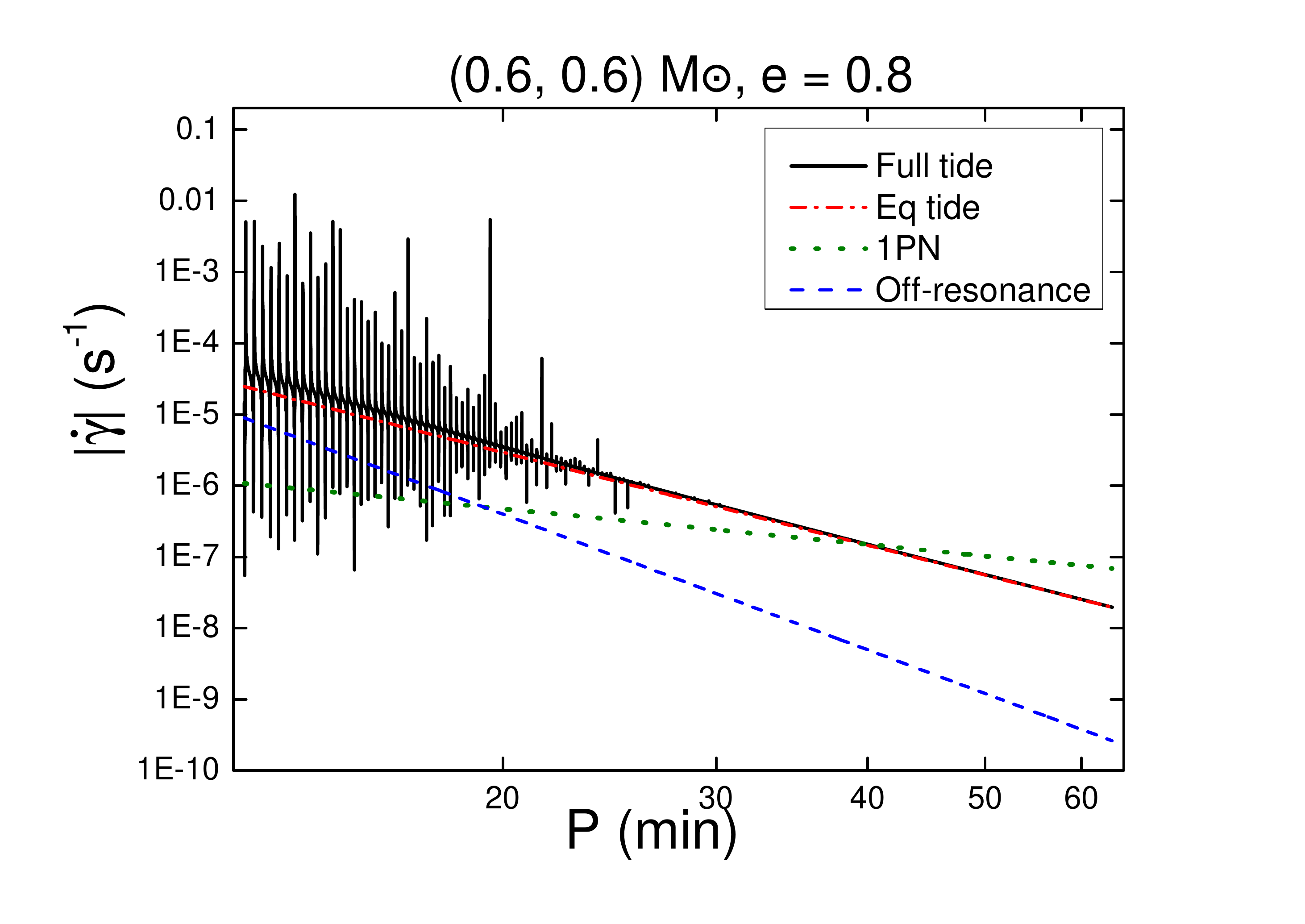}
    \caption{\label{fig:prec_full} The absolute value of the precession rate of DWDs with different masses and eccentricities. The upper left panel has (0.2, 0.6)~$M_\odot$ and $e = $0.5, and the others have (0.6, 0.6)~$M_\odot$ and $e = $0.1, 0.5, 0.8. The full tidal contribution is given in solid black lines. The individual factors of precession include the  equilibrium tide (dash-dotted red), 1PN (dotted green), and off-resonant dynamical tide (dashed blue). The smallest $P$ corresponds to the separations where the WDs fill the Roche lobe at pericenter.}
\end{figure*}

\subsection{\label{SSec:Precession_Formulation} The precession formula}
A useful way to quantify the amount of precession caused by the tidal interaction is to employ the method of osculating orbits.
At any given moment, the orbit can be described by a set of orbital elements of a Keplerian orbit with the same instantaneous position and velocity. This allows us to express the perturbed orbit with time-changing orbital elements caused by the tidal interaction. In particular, we employ the precession rate equation from \cite{doi:10.1119/1.10237}
\begin{align}
    \frac{d \gamma}{d t} =& \sqrt{\frac{a(1-e^2)}{GM e^2}}\Bigg[ - a_r \cos{\Phi} \nonumber \\
    & + a_\phi \sin{\Phi} \left(\frac{2+e\cos{\Phi}}{1+e\cos{\Phi}}\right)   \Bigg]. \label{eq:prec_orb_avg}
\end{align}
Here, $\gamma$ is the argument of pericenter, $a$ is the semi-major axis, $e$ is the eccentricity, $a_r$ and $a_\phi$ are the radial and tangential components of the tidal acceleration given in Eq.~\eqref{eq:tide_acc} respectively. The secular change of $\gamma$ over one complete radial orbit is denoted by $\Delta \gamma$. 

We express the driven part of the solution of Eq.~\eqref{eq:mode_amp_eom} as a Fourier series in the orbital frequency $\Omega = \sqrt{GM/a^3}$. Together with the introduction of the Hansen coefficients, $X_k^{l,m}(e)$, defined by
\begin{align}
    \left(\frac{a}{D}\right)^{l+1} e^{-i m \Phi} = \sum_k X_k^{l,m}(e) e^{-i k \Omega t},
\end{align}
we have
\begin{align}
    \Delta \gamma =& \sum_{\alpha l m} \left(W_{lm} I_{\alpha l m}\right)^2\left(\frac{R_1}{a}\right)^{l+1} \nonumber \\
    &\times \sum_{k} \left(\frac{\omega_\alpha^2}{\omega_\alpha^2-k^2\Omega^2}\right) X_k^{l,m} A_k^{l,m}, \label{eq:tide_prec_series}
\end{align}
where $k$ is the Fourier series index of the Hansen coefficients, and
\begin{align}
    A_k^{l,m} =& \frac{2\pi }{e\sqrt{1-e^2}} \Bigg\{\left(\frac{l+1}{2}-m\right)X_k^{l, m-1} \nonumber \\
    &+ \left(\frac{l+1}{2}+m\right)X_k^{l, m+1} \nonumber\\ 
    &+ \frac{e}{4}\Big[\left(l+1-m\right)X_k^{l, m-2}+2(l+1)X_k^{l, m}\nonumber\\
    &+\left(l+1+m\right)X_k^{l, m+2}\Big]\Bigg\}. \label{eq:tide_Almk}
\end{align}
The full tidal contribution to $\Delta\gamma$ is separated into the equilibrium component and the dynamical component, $\Delta \gamma = \Delta \gamma_\text{eq} + \Delta \gamma_\text{dyn}$, by decomposing the factor in the mode amplitude as
\begin{align}
\left(\frac{\omega_\alpha^2}{\omega_\alpha^2-k^2\Omega^2}\right) = 1+\left(\frac{k^2\Omega^2}{\omega_\alpha^2-k^2\Omega^2}\right),
\end{align}
where the first term denotes the equilibrium component and the second one is the dynamical component.

For $k\Omega$ not near $\omega_\alpha$, we can take the leading order of $\Delta \gamma_\text{dyn}$, given by 
\begin{align}
    \Delta \gamma_\text{off-res} =& \sum_{\alpha l m}\frac{m_2}{m_1} \left(W_{lm} I_{\alpha l m}\right)^2 \left(\frac{R_1}{a}\right)^{2l+1} \nonumber \\
    &\times \left(\frac{\Omega}{\omega_\alpha}\right)^2 \sum_{k} k^{2} \left(X^{l,m}_{k} A^{l,m}_{k} \right).\label{eq:dyn_tide_expd}
\end{align}
Here, we define $\Delta \gamma_\text{off-res}$ as the off-resonant approximation of $\Delta \gamma_\text{dyn}$, which has a simple power law dependence on the orbital frequency as $\Omega^{16/3}$ obtained by applying the Kepler's third law of orbital motion on $1/a^5$.

\subsection{\label{SSec:prec_factor_comparison}Comparison with other factors of precession}

It has been shown in \cite{PhysRevLett.100.041102} that the equilibrium tide contribution to the precession rate dominates at intermediate separations, while the 1PN contribution becomes important at larger separations. 
The equilibrium tide precession in the quadrupolar order, $\Delta \gamma_{\text{eq}}$, can be written in a simple form (see, e.g., \cite{poisson_will_2014})
\begin{align}
    \Delta \gamma_{\text{eq}} =& 30\pi k_1\frac{m_2}{m_1}\left[\frac{R_1}{a (1-e^2)}\right]^5 \left(1+\frac{3}{2}e^2+\frac{1}{8}e^4\right), \label{eq:prec_orb_avg_eq_tide}
\end{align}
where $k_1$ is the $l=2$ tidal Love number of WD 1. This quantity depends on the mass distribution of the WD and is related to the overlap integral by 
\begin{align}
    k_1 =& \sum_{\substack{n\\
                  |m|\le2}} \left(W_{2m} I_{\alpha 2 m}\right)^2\nonumber \\
                  \approx& \frac{4\pi}{5} I_{\text{\textit{f}-mode}}^2, \label{eq:tidal_def_sum}
\end{align}
where $I_{\text{\textit{f}-mode}}$ is the overlap integral of the $l=2$ $f$-mode. The 1PN precession is given by
\begin{align}
    \Delta\gamma_\text{1PN} = 6\pi \frac{GM}{c^2 a (1-e^2)}.\label{eq:1PN_prec}
\end{align}
Both precession effects are positive, meaning they are in the same direction as the orbital motion.

In Fig.~\ref{fig:prec_full}, we compare the full tidal contribution, Eq.~\eqref{eq:tide_prec_series}, for the precession rate averaged over one orbit, $\dot{\gamma}\equiv \Delta \gamma / P$, with Eqs.~\eqref{eq:prec_orb_avg_eq_tide} and \eqref{eq:1PN_prec} for a (0.2, 0.6)~$M_\odot$ DWD system with $e = 0.5$ and (0.6, 0.6)~$M_\odot$ systems with $e = 0.1, 0.5, 0.8$ at different orbital periods, denoted by $P$. We also show the off-resonant approximation of the dynamical tide from Eq.~\eqref{eq:dyn_tide_expd}. 

The full tidal contribution is calculated using Eq.~\eqref{eq:tide_prec_series}, summing up $k$ from $-k_{max}$ to $k_{max}$, where the value of $k_{max}$ is chosen to be 
\begin{align}
    k_{max} = 16 k_\mathrm{peri} = 16 \; \text{Int}\left[\frac{(1+e)^2}{(1-e^2)^{3/2}}\right],
\end{align}
where Int[...] means taking the nearest integer of the argument, and $k_\mathrm{peri}$ represents the harmonic corresponding to the motion closest to pericenter passage. The result shows a resonant response when the frequency of a harmonic comes close to the $f$-mode frequency, causing a series of narrow peaks at different $P$. Unlike the equilibrium tide and the 1PN effect, the precession caused by resonance can be orders of magnitude larger than the other effects and can be negative for orbits just inside the resonance. As the eccentricity increases, the resonance peaks also become more significant and cover a larger range of $P$. Note that even though the peaks appear only in the part with small $P$ in Fig.~\ref{fig:prec_full}, they are evenly distributed along the horizontal axis. For peaks at larger $P$, it requires a more refined grid to show them in the plot due to the decreased width of resonance. We shall discuss these resonant effects on orbital precession in Sec.~\ref{ssec:prec_at_resonance}.

The 1PN, equilibrium tide, and the off-resonant part of the dynamical tide all contribute to positive precession of the orbit, and have power law dependence of $P^{-5/3}$, $P^{-13/3}$ and $P^{-19/3}$ respectively. As a result, the 1PN effect dominates at large separations, while the tidal contributions, mainly the equilibrium tide, become the dominant effects at small separations. The off-resonant contribution from the dynamical tide is a relatively small effect but increases rapidly when the system is close to Roche-lobe filling separation, becoming comparable to the equilibrium tide for extremely close orbits. Comparing the (0.2, 0.6) and (0.6, 0.6)~$M_\odot$ cases, an increase in mass causes the tidal effect on $\dot{\gamma}$ to decrease for orbits with the same period and eccentricities. However, the more massive systems can also have closer orbits before filling the Roche lobe, which results in a larger maximum tidal effect.

\subsection{\label{SSec:prec_off_res} Off-resonant contribution to precession}

\begin{figure}
    \centering
    \includegraphics[width=8.6cm]{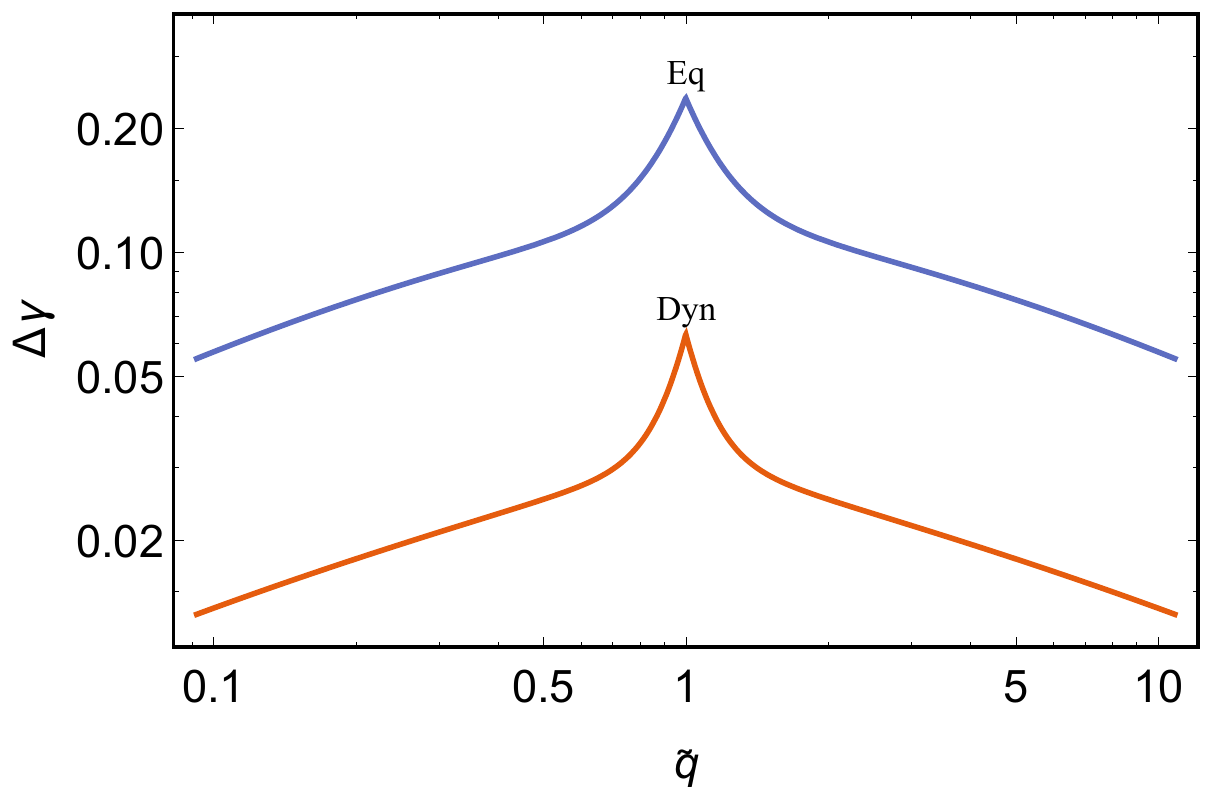}
    \caption{\label{fig:prec_analysis_q} The precession angle caused by the different components of the tide of DWDs at $e\rightarrow0$ limit with pericenter separation $r_p = a_{RL}$ with different mass ratio $\tilde{q} = m_1/m_2$. We only include the off-resonant contribution for dynamical tides. 
    }
\end{figure}

While the effect of resonance is enormous, it only occurs in a narrow region of the frequency space. To quantify the maximum potential of the off-resonant contribution from the dynamical tide, we explore its dependence on the orbital parameters at the $e\rightarrow0$ limit of Eq.~\eqref{eq:dyn_tide_expd} that contains a small number of terms:
\begin{align}
    \Delta \gamma_{\text{off-res}} =& 447.5 \left(\frac{m_2}{m_1}\right)\left(1+\frac{m_2}{m_1}\right) (I_{\alpha l m})^2\left(\frac{R_1}{a}\right)^8, \label{eq:prec_dyn_e0_limit}
\end{align}
where we have used the low eccentricity expansion of $X_k^{l,m}$ (see, e.g., \cite{Weinberg_2012}):
\begin{align}
    X_k^{l,m}(e) =& \delta_{k, m} + \frac{e}{2}[(l+1-2m)\delta_{k, m-1} \nonumber \\
    &+ (l+1+2m)\delta_{k, m+1}] + O(e^2), \label{eq:Hansen_small_e}
\end{align}
and we have assumed that $\omega_\alpha$ is proportional to $\sqrt{G m_1/R_1^3}$. We choose the proportionality constant such that $\omega_\alpha = 0.52~\text{s}^{-1}$ when $m_1 = 0.6~M_\odot$ and $R_1 = 8840$~km. The maximum possible effect from detached DWD is when $a$ equals the Roche-lobe filling separation, $a_{RL}$. We employ Eggleton's formula (\cite{1983ApJ...268..368E}) to approximate $a_{RL}$, which is a function of mass ratio $\tilde{q}=m_1/m_2$ and $R_1$. Equation~\eqref{eq:prec_dyn_e0_limit} can be further simplified by relating the WD radius to the mass, which we employ the approximate relation by \cite{1988ApJ...332..193V}. The overlap integral, $I_{\alpha l m}$, has a weak dependence on the mass and is set to 0.2. Hence, from Eq.~\eqref{eq:tidal_def_sum}, $k_1 \approx 0.1$.

The expressions in Eqs.~\eqref{eq:prec_orb_avg_eq_tide} and \eqref{eq:prec_dyn_e0_limit} are plotted in Fig.~\ref{fig:prec_analysis_q} for different $\tilde{q}$ including the contributions from both WDs. The precession effect of both the dynamical tide and equilibrium tide reaches a maximum at $\tilde{q} = 1$. We see that the off-resonant dynamical tide contributes to a maximum of about 20\% of the precession caused by the overall tidal effect.

As the DWD system approaches the Roche-lobe filling separation, the dynamical tide precession starts to deviate significantly from the off-resonant approximation due to the increase in the resonance width. In the following subsection, we focus on analysing the significance of these resonance peaks.

\subsection{\label{ssec:prec_at_resonance} The width of resonance}

As the pericenter separation, $r_p$, gets closer to $a_{RL}$, the harmonics near $k_\mathrm{peri}$, which have large Hansen coefficients, have frequencies closer to the $f$-mode frequency. As a result, we expect a stronger resonance effect when the system is close to Roche-lobe filling. To quantify the significance of such an effect, we calculate the width of the resonance where the contribution from a single mode dominates the overall precession due to tides. 

\begin{figure} [!tp]
    \centering
    \includegraphics[width=8.6cm]{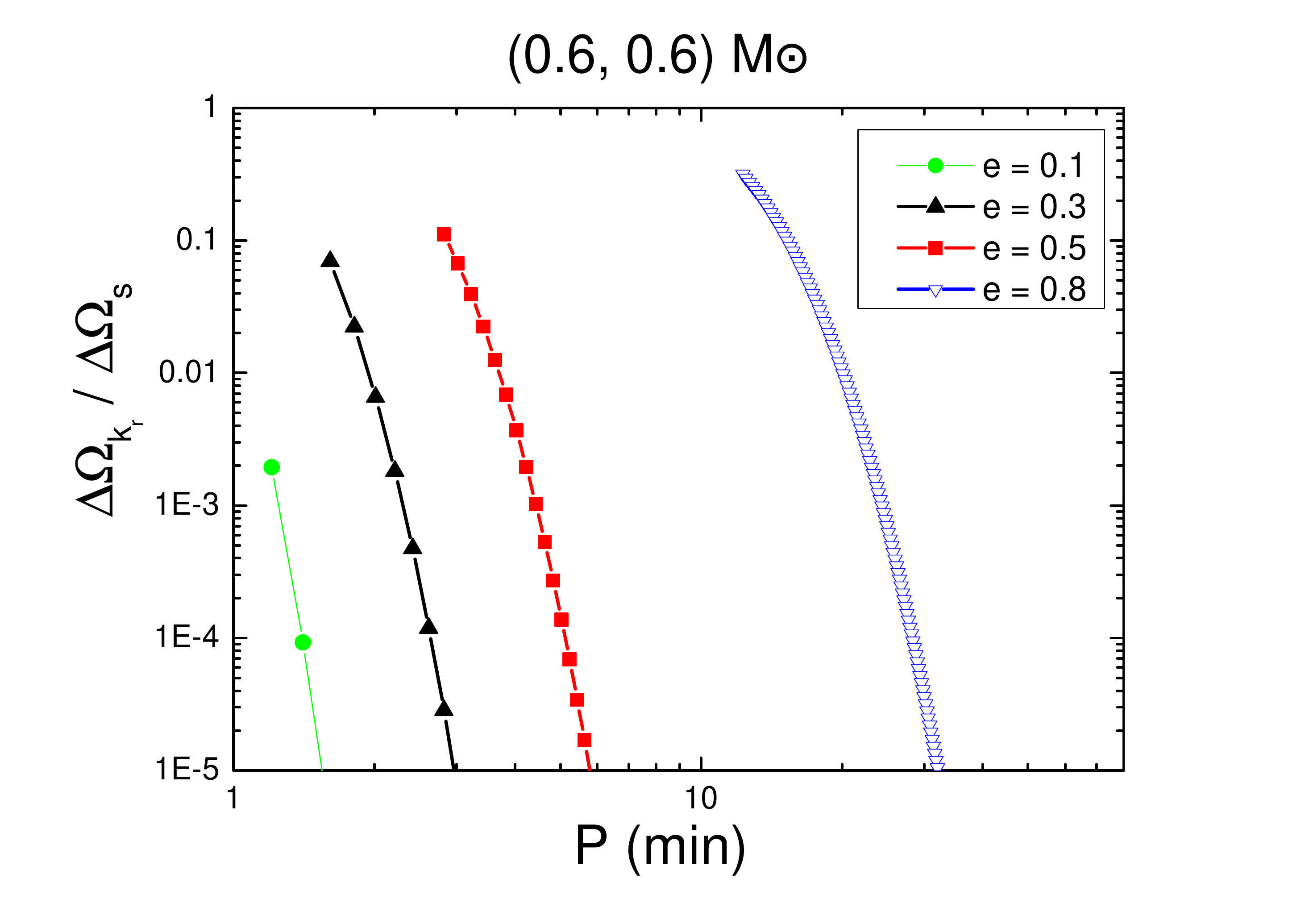}
    \caption{\label{fig:prec_res_width} The width-to-separation ratio of the resonance of (0.6, 0.6)~$M_\odot$ DWDs with different eccentricities. The smallest $P$ of each curve corresponds to the orbital period with a harmonic in resonance right before Roche-lobe filling.}
\end{figure}

We rearrange the summations in Eq.~\eqref{eq:tide_prec_series} to write it as a sum of contributions from each harmonic
\begin{align}
    \Delta \gamma = \sum_k \Delta \gamma_k.
\end{align}
At any orbital frequency, there is one harmonic closest to the $f$-mode frequency, which the harmonic order is denoted by $k_r$. The resonance width, $\Delta \Omega_{k_r}$, is defined to be the range of frequency where the magnitude of a single mode contribution to the precession exceeds the overall equilibrium tide contribution, i.e., the region with condition $|\Delta \gamma_{k_r}| \ge \Delta \gamma_\text{eq}$. The width-to-separation ratio of the resonance, $\Delta \Omega_{k_r} / \Delta \Omega_s$, of DWDs with different eccentricities are shown in Fig.~\ref{fig:prec_res_width}. Here, $\Delta \Omega_s$ is the frequency difference between subsequent resonance, taken to be $\omega_\alpha/(k_r-1)$. It shows that within the more eccentric systems, the width of resonance takes about 10\% of the separation between harmonics when the system is close to Roche-lobe filling separations. This ratio decays very rapidly as the orbital period increases. For systems with small eccentricities, only a few resonance peaks with the smallest $P$ have significant width. This ratio increases with eccentricities since $\Delta \Omega_s$ scales as $(1-e)^{3}$, while $\Delta \Omega_{k_r}$ stays within the same order of magnitude at a fixed pericenter distance for different eccentricities.

\subsection{\label{ssec:prec_with_spin} The effect of spin}

So far we have ignored the effect of rotation of the WD on the dynamical tide. One reason is the complexity when we include the Coriolis force, which changes the spectrum of eigenmodes \cite{10.1093/mnras/182.3.423}. Also, the angular dependence of the modes can no longer be expressed simply by a single spherical harmonic \cite{1989nos..book.....U}. Here we want to estimate how the Coriolis force affects the results without including the full details. If we treat the effect of the Coriolis force as a small perturbation to the mode frequency and ignore the changes to $I_{\alpha l m}$, the mode amplitudes in the inertial frame have a similar expression as those in the non-rotating case:
\begin{align}
    q_{\alpha}^{(k)} = \frac{\sigma_{\alpha}}{\sigma_{\alpha} -k\Omega + m\Omega_\text{s}} U_\alpha^{(k)}, \label{eq:mode_amp_spin}
\end{align}
where $\Omega_\text{s}$ is the spin rate of the WD,  $\sigma_{\alpha}$ is the mode frequency including the correction due to Coriolis force observed in the rotating frame, which is given in the leading order of the spin rate by (see e.g.,  \cite{1989nos..book.....U, Lai_1997, Pnigouras:2022zpx})
\begin{align}
    \sigma_{\alpha} = \omega_\alpha^{(0)} - m\Omega_\text{s} C_{n l},
\end{align}
where $\omega_\alpha^{(0)}$ is the eigenfrequency of the non-rotating WD, i.e., the quantity denoted by $\omega_\alpha$ in Sec.~\ref{Sec:dyn_tide_formulation} and the rest of the paper where the WD's rotation is not included, and the coefficient $C_{n l}$ depends on the mode eigenfunctions of the non-rotating WD. The explicit form of this coefficient can be found in e.g., \cite{1989nos..book.....U, Lai_1997}. In this subsection, we take $C_{n l} = 1/l$ for $f$-modes, which can be shown to be the exact result for an incompressible star and is a good approximation for the WD models.

\begin{figure} [!tp]
    \centering
    \includegraphics[width=8.6cm]{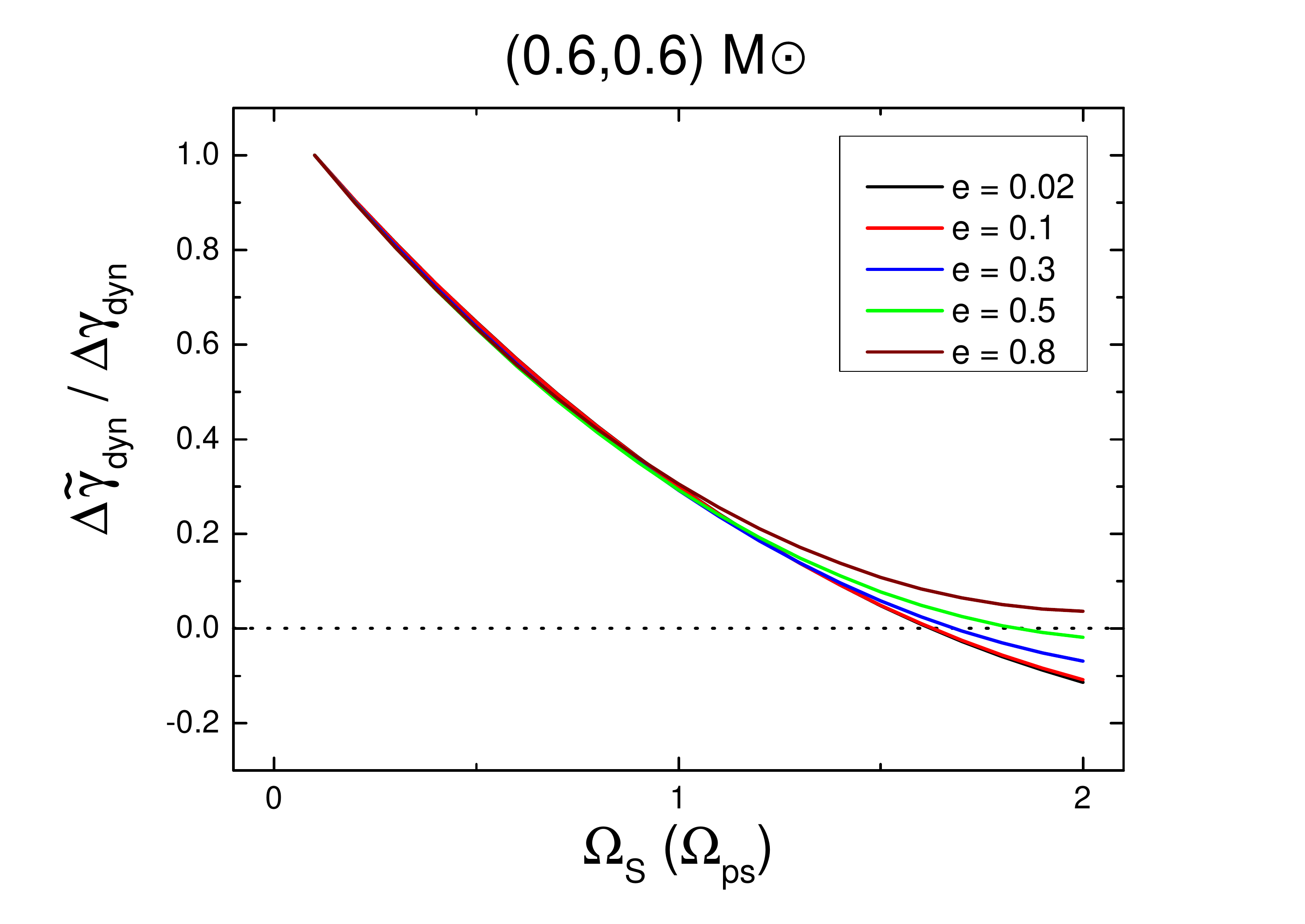}
    \caption{\label{fig:prec_vary_spin} 
  The normalized precession angle of the (0.6, 0.6)~$M_\odot$ DWDs including the correction on the mode frequency due to the Coriolis effect at different spin rates. The WDs are in an orbit with $r_p=2~a_{RL}$. A horizontal dotted line is used to indicate $\Delta\tilde{\gamma}_{\text{dyn}} = 0$.}
\end{figure}

Including this change on the mode frequencies and amplitudes, we have the precession angle $\Delta \tilde{\gamma}_\text{dyn}$ with spin correction given by
\begin{align}
    \Delta \tilde{\gamma}_\text{dyn} =& \frac{m_2}{m_1} \sum_{\alpha l m } \sum_{k} \left(W_{lm} I_{\alpha l m}\right)^2  X_k^{l,m}A_k^{l,m} \nonumber \\
    &\times \left[\frac{(k\Omega-m\Omega_\text{s})(k\Omega-m\Omega_\text{s} + m C_{nl} \Omega_\text{s})}{\omega_\alpha^{(0) 2}-(k\Omega-m\Omega_\text{s} + m C_{nl} \Omega_\text{s})^2}\right]. \label{eq:dyn_tide_prec_spin}
\end{align}
This equation shows that the resonance frequency of the $f$-mode is split into the modified mode frequencies $\omega_\alpha^{(0)} + m(1-C_{nl})\Omega_s$.

To estimate the effect of a range of rotation rates, we vary the spin from 0 to twice the pseudo-synchronous value, $\Omega_\text{ps}$, determined using the weak friction model by Hut \cite{1981A&A....99..126H},
\begin{align}
    \Omega_\text{ps} = \frac{\Omega}{(1 - e^2)^{3/2}} \left(\frac{16+120e^2+90e^4+5e^6}{16+48e^2+6e^4}\right). \label{eq:pseudosync}
\end{align}
This equation describes the spin rate of the WD such that there is no net torque from the equilibrium tide onto the orbit. The more general tidal synchronization problem involving dynamical tide involves the damping mechanisms of the eigenmodes and is not considered here.

In Fig.~\ref{fig:prec_vary_spin}, we show the change of $\Delta \tilde{\gamma}_\text{dyn}$ normalized by the precession angle of the non-rotating WD, $\Delta \gamma_\text{dyn}$, when the spin rate of both of the WDs increases from 0 to 2~$\Omega_\text{ps}$, with the pericenter separation fixed at 2~$a_{RL}$. As the spin rate increases, $\Delta \tilde{\gamma}_\text{dyn}$ decreases and crosses zero at some value larger than $\Omega_\text{ps}$ depending on the eccentricity. It shows that if the WD has a high spin rate, the dynamical tide precession is suppressed or even becomes negative. As the eccentricity increases, the $\Omega_\text{s}$ at which $\Delta \tilde{\gamma}_\text{dyn}$  crosses zero increases until it stays above zero when $e\gtrsim$ 0.8.

The precession angle of the lower eccentricity orbits has a steeper dependence on $\Omega_\text{s}$. We illustrate a specific example with the $e\rightarrow0$ limit of Eq.~\eqref{eq:dyn_tide_prec_spin} that contains a small number of terms, which we explicitly write down as
\begin{widetext}
\begin{align}
    \Delta \tilde{\gamma}_\text{dyn} =& \frac{m_2}{m_1} \sum_{\alpha} \left(\pi I_{\alpha l m}\right)^2\left(\frac{R_1}{a}\right)^{5} \Bigg[\left(\frac{9}{5}\right)\frac{\Omega^2}{\omega_\alpha^{(0) 2}-\Omega^2}+\left(\frac{3}{10}\right)\frac{(\Omega-2\Omega_\text{s})[\Omega-2(1-C_{nl}) \Omega_\text{s}]}{\omega_\alpha^{(0) 2}-[\Omega-2(1-C_{nl}) \Omega_\text{s}]^2} \nonumber \\
    &-\frac{24(\Omega-\Omega_\text{s})[\Omega- (1-C_{nl}) \Omega_\text{s}]}{\omega_\alpha^{(0) 2}-4[\Omega- (1-C_{nl}) \Omega_\text{s}]^2} +\left(\frac{147}{10}\right)\frac{(3\Omega-2\Omega_\text{s})[3\Omega-2 (1-C_{nl})\Omega_\text{s}]}{\omega_\alpha^{(0) 2}-[3\Omega-2 (1-C_{nl})\Omega_\text{s}]^2}\Bigg]. \label{eq:dyn_tide_prec_spin_e0}
\end{align}
\end{widetext}
The value of $\Omega_\text{s}$ which causes $\Delta \tilde{\gamma}_\text{dyn}$ to become zero can be solved analytically in the off-resonant approximation. The result is independent of $\omega_\alpha^{(0)}$ and is found to be $\Omega_\text{s} = (81-\sqrt{1041})/30~\Omega_\text{ps} \approx 1.62~\Omega_\text{ps}$, where we used $\Omega = \Omega_\mathrm{ps}$ when $e=0$ from Eq.~\eqref{eq:pseudosync}.

The centrifugal force from the rotating WDs deforms the stars and causes extra precession. Willems \textit{et al.} \cite{PhysRevLett.100.041102} compared this effect with that of the 1PN and equilibrium tide. They showed that for systems close to pseudo-synchronous, it gives a precession rate with the same power law dependence on orbital separation as the equilibrium tide, but several times smaller in size. Therefore, while the Coriolis force suppresses the dynamical part of the tide, the centrifugal force gives an extra precession that enhances the equilibrium tide effect by an amount depending on the rotation rate.

\section{\label{Sec:Precession_on_waveform} The effect of precession on gravitational wave detection}

\subsection{\label{SSec:Precession_detectability} The parameter space affected by the dynamical tide}

\begin{figure*}
    \includegraphics[width=8.6cm]{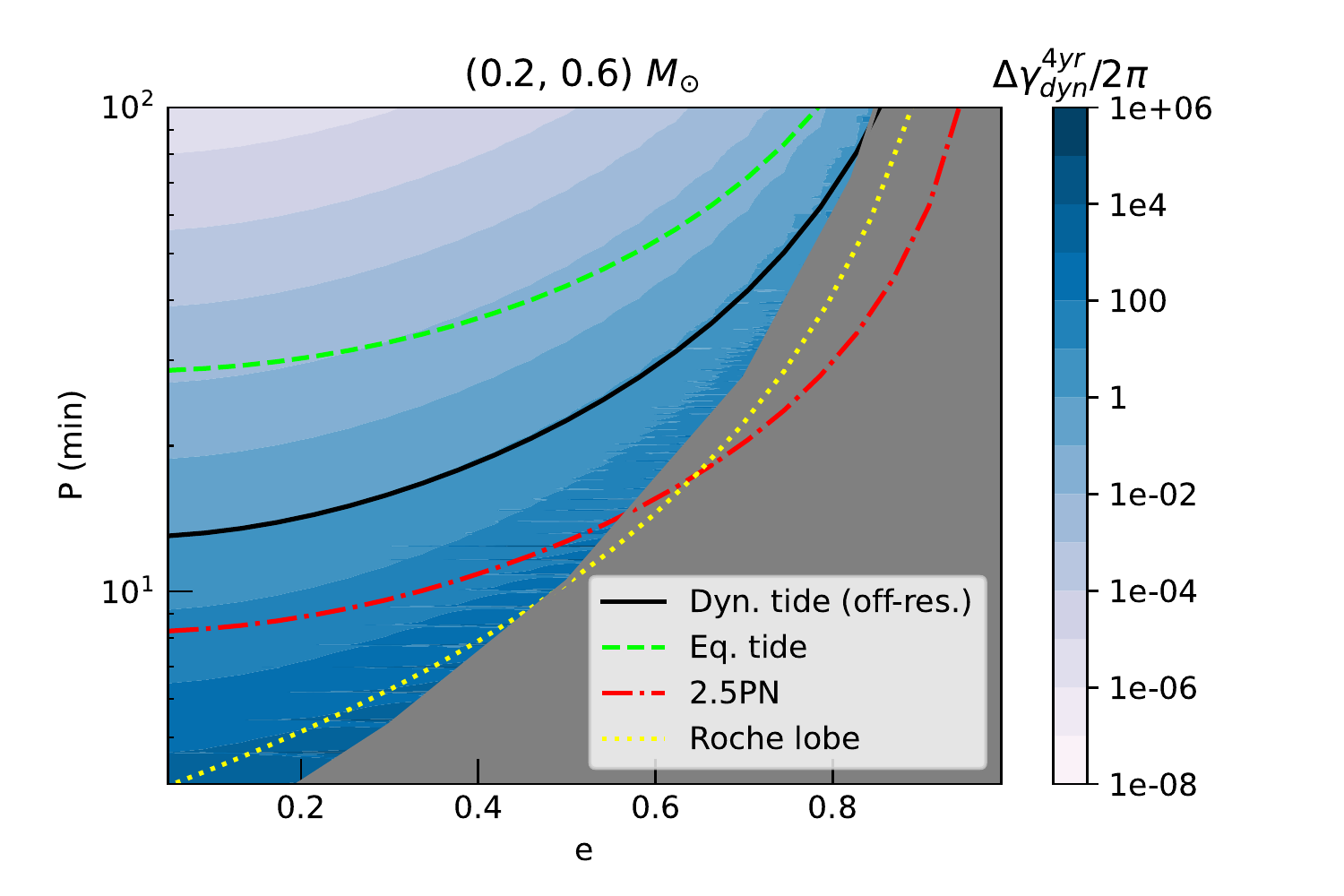}
    \includegraphics[width=8.6cm]{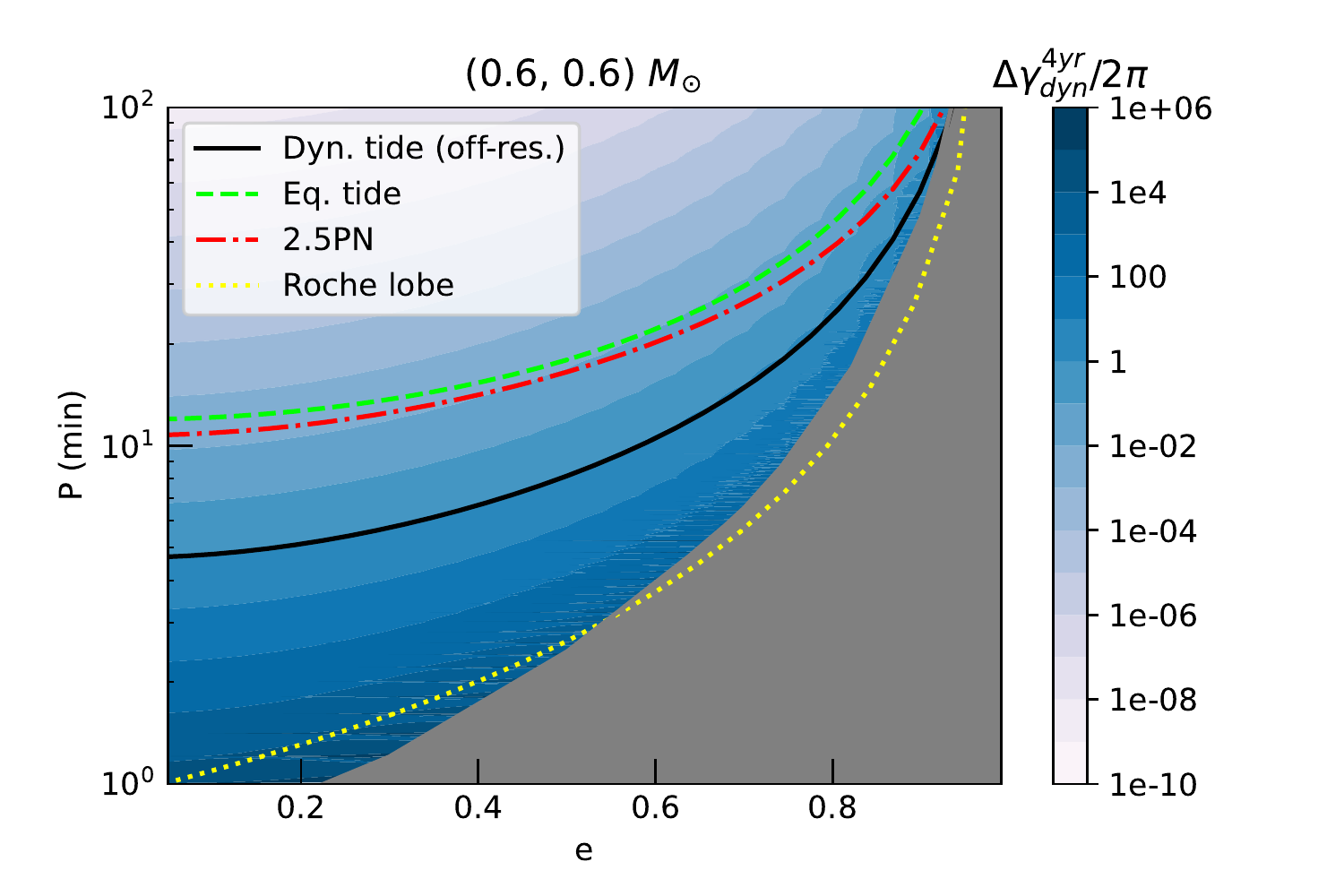}
 \caption{\label{fig:LISA_resolution} The total phase shift caused by the dynamical tide precession (including both resonant and off-resonant contributions) divided by $2\pi$ in a 4-year observation, $\Delta \gamma_\text{dyn}^\text{4yr}/(2\pi)$ (color contours). The maximum $P$ for the frequency change caused by the precession from the off-resonant dynamical tide (black solid line), equilibrium tide (green dashed line), and the chirping of the 2.5PN orbital decay (red dash-dotted line) to be within the resolution of LISA are also shown. The orbital period corresponding to Roche-lobe filling separation is indicated with the yellow dotted line. The DWD systems have masses (0.2, 0.6)~$M_\odot$ (left panel) and (0.6, 0.6)~$M_\odot$ (right panel). The grey region indicates the parameter space for which the orbit performs chaotic motion.  }
 \end{figure*}
 
The detection of the dynamical tide effect is limited by the resolution of LISA. A shift in phase over the observation period is resolvable if it exceeds $2\pi$. In Fig.~\ref{fig:LISA_resolution}, we show the total phase shift (in units of $2\pi$) caused by dynamical tide precession over 4 years within the ($e, P$) parameter space for a (0.2, 0.6)~$M_\odot$ and a (0.6, 0.6)~$M_\odot$ DWD system. The phase shift increases towards the large $e$ and small $P$ region. The horizontal strips appearing in the high eccentricity region correspond to the large phase shift due to resonance between a certain harmonic and the $f$-mode. These resonances provide a phase shift larger than $2\pi$ even for relatively large $P$. Note that due to the limited resolution of the plot, the resonances appear to vanish at lower eccentricity as the resonance width decreases, even though they are expected to extend all the way to the low eccentricity region.

The maximum orbital period required for the precession caused by the equilibrium tide and the off-resonant dynamical tide to remain resolvable by LISA are represented by the green dashed line and the black solid line respectively. Similarly, the leading order effect on the frequency shift due to GW emission, i.e., the 2.5PN effect (given in \cite{PhysRev.131.435}), is also subject to this limit and is shown with the red dash-dotted line. In the (0.2, 0.6)~$M_\odot$ system, the off-resonant dynamical tide effect is resolvable for a larger range of $P$ than the 2.5PN effect, while it is the opposite for the (0.6, 0.6)~$M_\odot$ system. In both cases, the equilibrium tide effect has the largest resolvable range for all eccentricities.

The parameter space within the grey region at high eccentricities corresponds to chaotic evolution of the orbit. This phenomenon is caused by dynamical tide in highly eccentric binaries, first studied in \cite{1995ApJ...450..732M} (see also \cite{2001MNRAS.321..398M, 2004MNRAS.347..437I, 2018MNRAS.476..482V}). Ivanov and Papaloizou \cite{2004MNRAS.347..437I} illustrate that such a chaotic behaviour due to the secular accumulation of mode energy over many pericenter passages can be understood as a kind of stochastic instability, as the mode amplitude receives a phase change at each passage which can be approximated as a uniformly distributed random variable. Here, we employ the result by \cite{2004MNRAS.347..437I, 2018MNRAS.476..482V} to approximately map out the chaotic region for the DWD system, which is written as $|\omega_\alpha \Delta P| \geq 1$, with $\Delta P$ being the change in orbital period over one orbit due to the tide. In the case of (0.2, 0.6)~$M_\odot$, we consider only the $|\omega_\alpha \Delta P|$ of the 0.2~$M_\odot$ WD, as it is expected to provide the dominant tidal effect inside the binary.

\begin{figure}
\includegraphics[width=8.6cm]{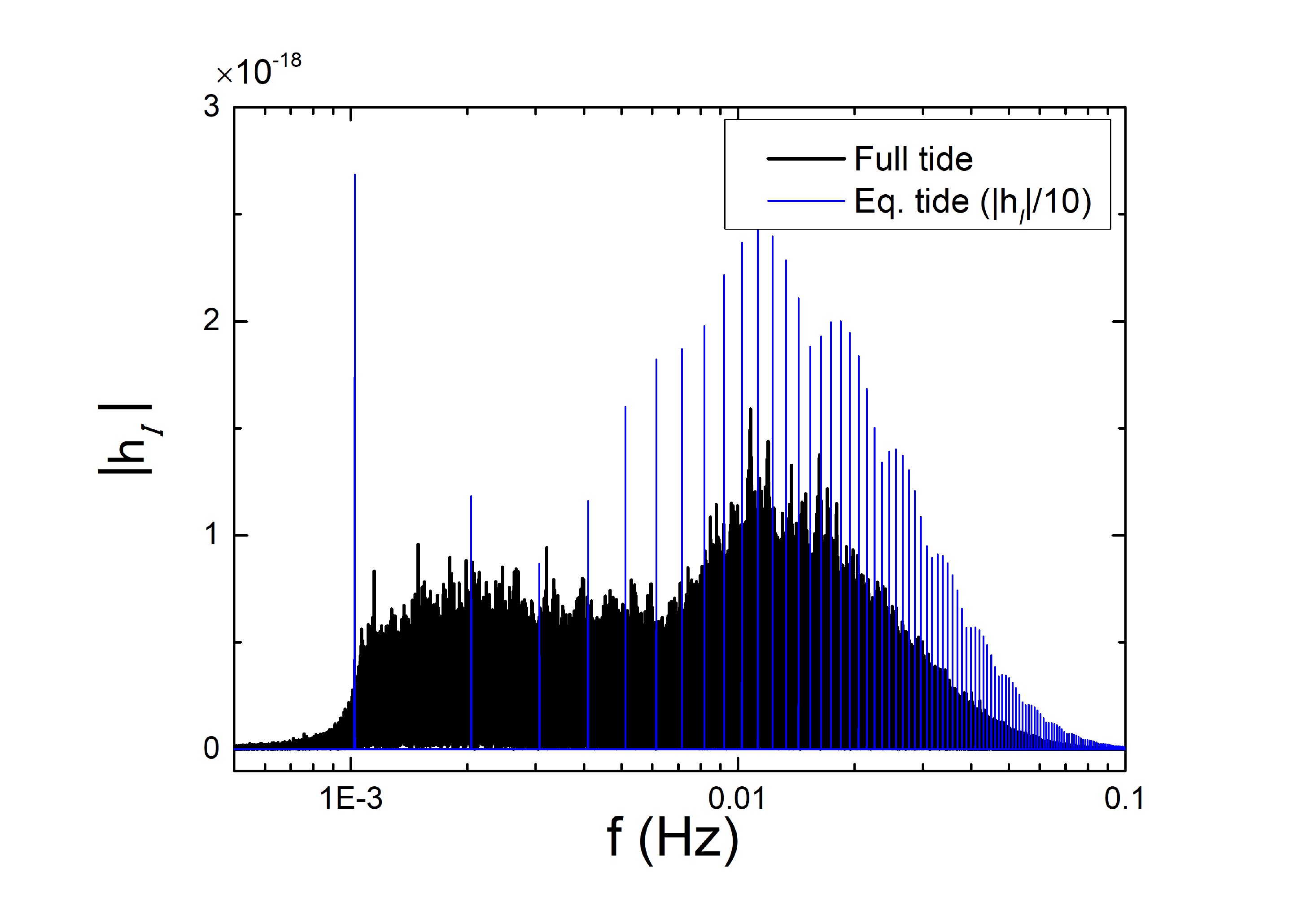}
\caption{\label{fig:chaos_h1} Frequency domain waveform amplitude of a (0.6, 0.6)~$M_\odot$ DWD system inside the chaotic regime obtained by numerically integrating the orbital equation of motion Eq.~\eqref{eq:eom} for a duration of 0.25 years. The initial eccentricity and pericenter distance are set to be 0.82 and 1.2~$a_{RL}$ (P = 14.7 min) respectively (within the grey region in Fig.~\ref{fig:LISA_resolution}), and the distance is at 10~kpc. The waveform including only equilibrium tide is shown in blue lines and is rescaled by a factor of 1/10.}
\end{figure}

This region of chaotic behaviour limits the detectability of the GW signal itself in the highly eccentric small separation regions, as the waveform is no longer predictable. In Fig.~\ref{fig:chaos_h1}, we show the frequency domain strain signal $|h_I|$ (Eq.~\eqref{eq:h_I}, see also Appendix~\ref{SSec:Numerical waveform}) of a system within the grey region of the (0.6, 0.6)~$M_\odot$ case due to the influence of full tide and equilibrium tide respectively. The initial conditions are chosen such that the pericenter separation is 1.2~$a_{RL}$, and the osculating Keplerian orbit has an eccentricity of 0.82. The distance from the source is set as 10~kpc. We see that the waveform inside the chaotic region has a spread of power along the frequency domain, instead of concentrating in the vicinity of each harmonic. This behaviour completely alters the waveform, making it impossible to detect with template matching. Note that even though the waveform becomes chaotic, it still has larger amplitudes in the vicinity of the harmonics corresponding to the angular speed near the pericenter, which is similar to the periodic waveform without dynamical tides.

The orbital period corresponding to Roche-lobe filling separation is indicated with the yellow dotted line in Fig.~\ref{fig:LISA_resolution}, setting the lower bound of the orbital period of detached DWDs. This line crosses the chaotic boundary with the chaotic region covering a larger $P$ at $e \gtrsim 0.5$. The crescent shape region in between the black solid curve and the lower bounds set by the Roche-lobe filling separation or the chaotic boundary, together with the resonant regions, represent the parameter space where the dynamical tide is resolvable by LISA.

Within these regions, the dynamical tide can have an impact on the waveform analysis, which we discuss in detail in the next subsection. However, it still requires a detailed Fisher analysis in order to quantify the actual measurability in parameter estimation. We shall leave it for future work.

\subsection{\label{SSec:prec_on_waveform_analysis}The effect of precession on waveform analysis}

The waveform from the eccentric DWDs consists of a superposition of nearly monochromatic signals at every orbital harmonic, with each of them split into a triplet with frequencies $k\Omega$, $k\Omega \pm 2\dot{\gamma}$ \cite{10.1093/mnras/274.1.115, PhysRevLett.100.041102}. The precession rate contains the combined effects of the tide and the 1PN effect. In the following, we consider whether it is possible to separate the dynamical tide from the other factors through waveform analysis.

A useful quantity that determines the distinguishability of two waveforms $h_1$ and $h_2$ with or without the dynamical tide effect is given by
\begin{align}
    || \delta h|| = \sqrt{\sum_{j=I,II}\langle h_{1, j} - h_{2, j} | h_{1, j} - h_{2, j} \rangle}, \label{eq:dh_LISA_def}
\end{align}
where $j=I,II$ is the interferometer index of LISA (\cite{PhysRevD.57.7089}). The inner product between two signals $a(t)$, $b(t)$ is defined by
\begin{align}
    \langle a| b \rangle = 4 \Re \int_0^\infty df \frac{\tilde{a}(f)\tilde{b}^*(f)}{S_n(f)},
\end{align}
where $\tilde{a}(f)$ and $\tilde{b}(f)$ are the Fourier transforms of $a(t)$ and $b(t)$. The symbol $\Re $ denotes taking the real part, and $S_n(f)$ is the noise spectral density of LISA, which we use the fitting formula in \cite{Robson_2019}. $ || \delta h||$ can be interpreted as the SNR of the difference between the signals. The two waveforms are said to be indistinguishable if $|| \delta h|| < 1$ \cite{PhysRevD.78.124020, PhysRevD.79.124033, MacDonald_2011}. Here, we employ the Peters and Mathews waveform model with precession for the plus and cross polarizations \cite{10.1093/mnras/274.1.115, PhysRev.131.435,PhysRevD.69.082005,PhysRevLett.100.041102}, whose explicit form is described in Appendix~\ref{SSec:PM_waveform}. The wave amplitude of this model at each harmonic follows that of a Keplerian eccentric orbit, which is found to be a good approximation based on our preliminary calculations using the numerical waveform in Appendix~\ref{SSec:Numerical waveform} for comparison. The antenna pattern acts to project the two polarizations into the detector strain signal, and is also given in Eq.~\eqref{eq:h_I}.

We assume the actual signal contains the full tidal contribution as well as the 1PN effect, and compute $||\delta h||$ by using a template without the dynamical tide component to estimate the difference caused by it. The eccentricity of the template is adjusted to match the precession rate with the signal. When the chirp is small, this is expected to maximize its match with the signal. 

\begin{figure}
    \includegraphics[width=8.6cm]{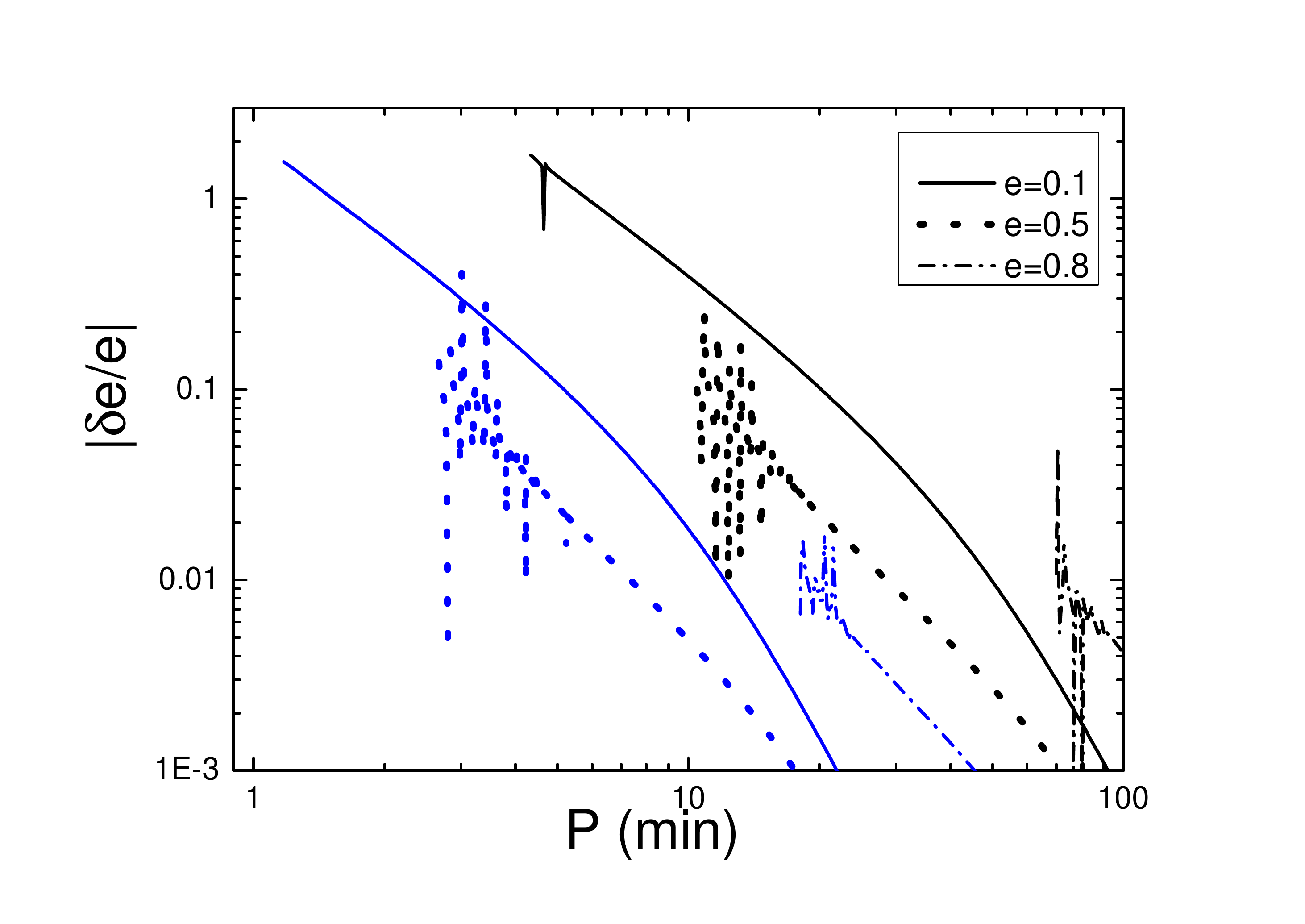}
    \caption{\label{fig:de_dyn_tide} The fractional difference of the eccentricities of the two different waveform models with and without the dynamical tide effect on the precession as a function of the orbital period $P$. Here, we fix the precession rate and orbital frequency in the two models to determine the fractional difference in $e$. The DWD systems have masses (0.2, 0.6)~$M_\odot$ (black lines) and (0.6, 0.6)~$M_\odot$ (blue) at different $P$. The pericenter separation of the smallest $P$ of each curve corresponds to the Roche-lobe filling separation, except for the $e=0.8$ case in which the closest separation is limited by the chaotic boundary in Fig.~\ref{fig:LISA_resolution}.}
\end{figure}

In Fig.~\ref{fig:de_dyn_tide}, we illustrate the effect of dynamical tide on the eccentricity measurement by showing the fractional difference of the eccentricities between the precession rate models with and without the dynamical tide. We assume the model with the full tidal contribution has eccentricity $e$, while the other model with the equilibrium tide contribution alone has eccentricity $e+\delta e$, such that the two models have the same precession rates and orbital frequencies. The result shows that the fractional shift in the eccentricity due to not including dynamical tide can be of the size of unity for orbits with low eccentricities at closest separations. At higher eccentricities, this shift is below several percent and is insignificant even for close orbits except at resonance. This is because the off-resonant dynamical tide precession effect decreases for orbits with larger eccentricities and fixed pericenter separations.
On the other hand, the near-resonant cases appear as narrow peaks and can have an order of magnitude difference in $|\delta e / e|$ from the off-resonant cases.

\begin{figure*}
    \includegraphics[width=8.6cm]{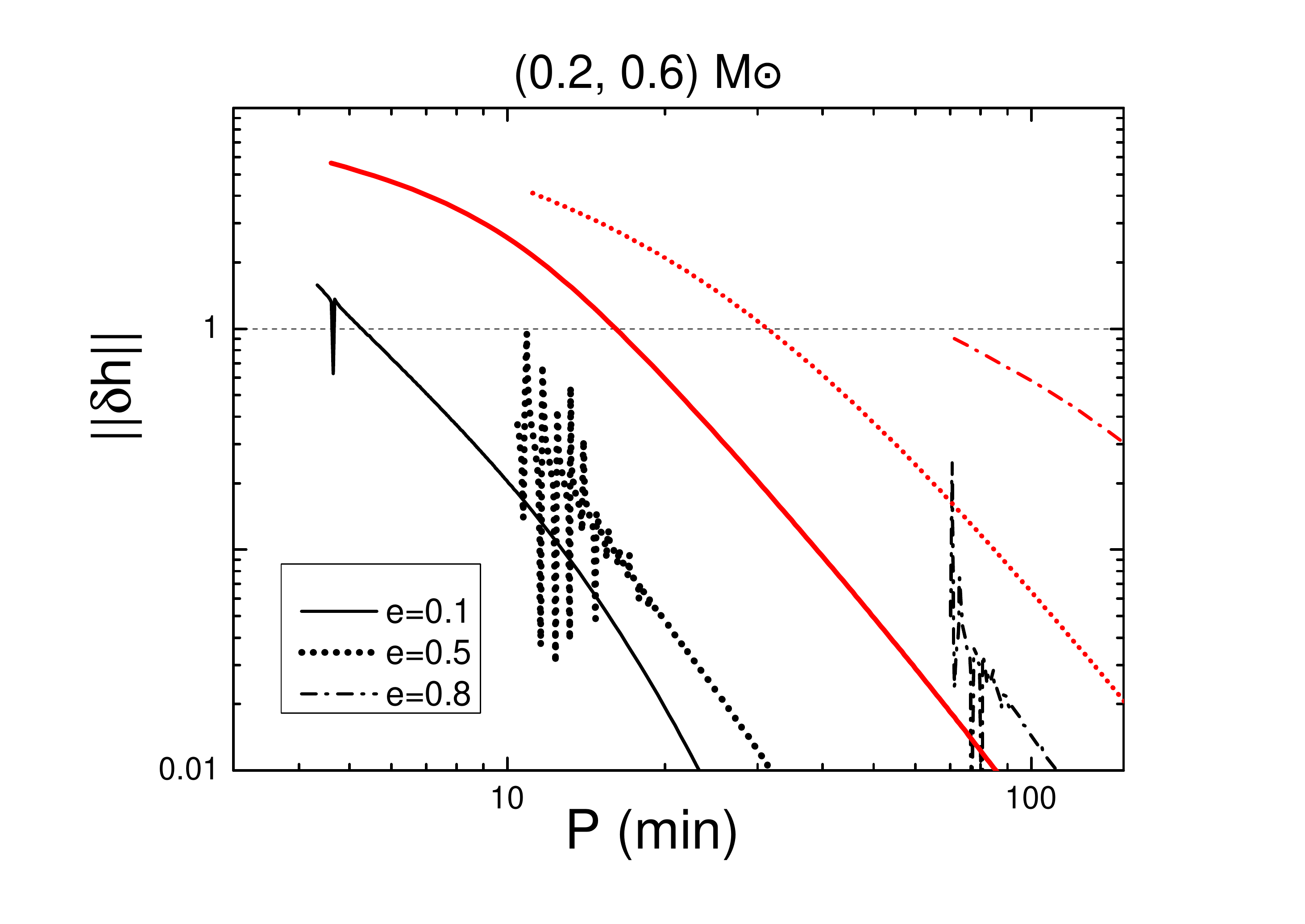}
    \includegraphics[width=8.6cm]{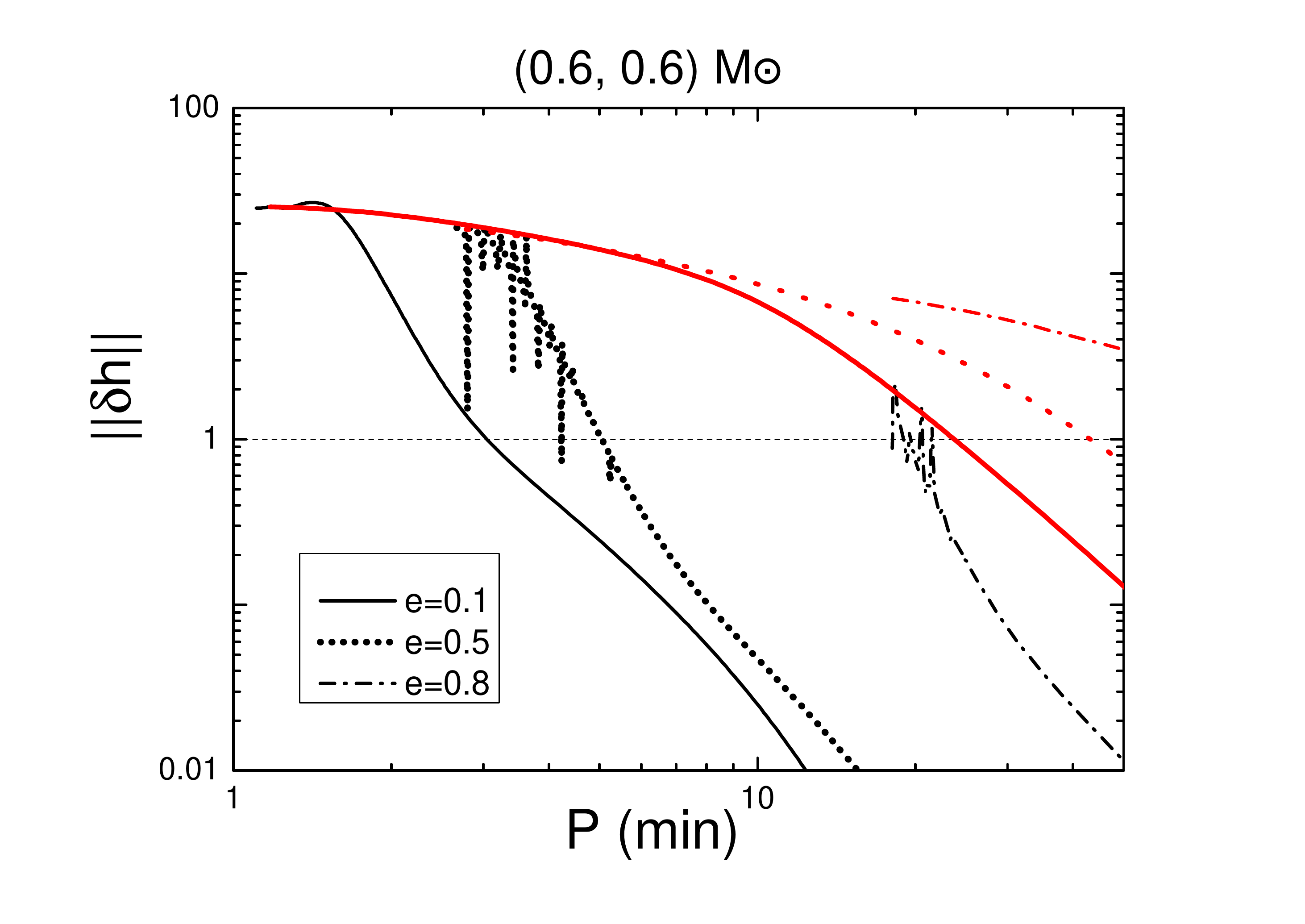}
    \caption{\label{fig:dh_dyn_tide} The quantity $||\delta h||$ of the (0.2, 0.6)~$M_\odot$ system (left panel) and the (0.6, 0.6)~$M_\odot$ system (right) in Fig.~\ref{fig:de_dyn_tide} are shown with black curves. The distance from the source, $d$, is taken to be 10~kpc. The observation time of the signal is set at 0.25yr. The horizontal dashed line corresponds to $||\delta h|| = 1$, the minimum value for the two signals to be distinguishable. The red curve shows  $\sqrt{||h_1||^2+||h_2||^2}$, the estimated value of $||\delta h||$ when the waveform with full tidal contribution is at resonance and the two signals are completely mismatched.}
\end{figure*}

We show the dependence of $||\delta h||$ on $P$ for the (0.2, 0.6) and the (0.6, 0.6)~$M_\odot$ DWD systems with different eccentricities in Fig.~\ref{fig:dh_dyn_tide}. For illustrative purposes, we assume the signal is observed for 0.25 years and the distance from the source is 10~kpc. The maximum SNR of the signal ranges from 0.5 to 2.5 for the (0.2, 0.6)~$M_\odot$ systems and 8 to 25 for the (0.6, 0.6)~$M_\odot$ systems depending on the eccentricity. We choose the observation period of 0.25 years for computational efficiency, and for a 4-year long signal, the SNR can be enhanced by a factor of 4. In both systems, $||\delta h||$ shows narrow peak features at small $P$ near resonance. The number of peaks appearing depends on the width of the resonance and the resolution of the plot. 

For orbits very close to resonance, the dynamical tide precession is much stronger than the other factors and cannot be replicated by choosing a waveform with an eccentricity within the reasonable range. The overlap between $h_1$ and $h_2$ will be small and the value of $||\delta h||$ can therefore be approximated as $\sqrt{||h_1||^2+||h_2||^2}$, where $||h_1||$ and $||h_2||$ are the SNRs of $h_1$ and $h_2$ with the same eccentricities respectively. We treat this as the maximum $||\delta h||$ at resonance, which are presented as red curves in Fig.~\ref{fig:dh_dyn_tide}. The resonance can cause the dynamical tide effect distinguishable from the other factors at a large $P$ ($>$20~min for small eccentricities with this parameter choice) given a signal with large SNR. Note that $\sqrt{||h_1||^2+||h_2||^2}$ is an approximate upper limit when the two waveforms have small match and is not necessarily always greater than $||\delta h||$. For the $e=0.1$ orbit of the $(0.6, 0.6)$~$M_\odot$ system at $P\approx 2$ min, $||\delta h||$ exceeds this value since $\sum_{j=I,II}\langle h_{1,j}|h_{2,j}\rangle < 0$. 

Focusing on the region outside resonance, the quantity $|| \delta h||$ is larger than 1 at small separations and approaches the maximum value in the case of (0.6, 0.6)~$M_\odot$ but is less than 1 for most separations in the (0.2, 0.6)~$M_\odot$ case, even though the fractional difference in eccentricities between $h_1$ and $h_2$ are similar (Fig.~\ref{fig:de_dyn_tide}). This different behaviour of $||\delta h||$ is mainly caused by the chirp of the signal. Since the 2.5PN effect also depends on eccentricity, measuring it helps resolve the degeneracy between the tidal effect and the eccentricity within the precession rate and allows us to identify the effect of dynamical tide\footnote{The 3.5PN effect on the phase \cite{10.1093/mnras/254.1.146} is found to cause less than 0.01$\%$ difference in $||\delta h||$ for the (0.6, 0.6)~$M_\odot$ at close separations, and hence is considered negligible in our calculations.}. This shows that LISA has the potential to identify the dynamical tide effect for high mass, eccentric DWD systems with low orbital periods, or systems with a higher harmonic close to resonance.

\section{\label{Sec:conclusion} Summary and Conclusion}

In this paper, we show that the dynamical tide has a strong influence on the eccentric DWD systems when they are close to the Roche-lobe filling separations, especially when the orbital motion resonates with the oscillation modes. At resonance, the dynamical tide causes a precession effect that can be orders of magnitude larger than that from equilibrium tide alone and can become negative in some frequency range as opposed to the equilibrium tide effect and the 1PN effect that only causes the pericenter to advance. The resonance is shown to take about 10\% of the frequency space within the more eccentric systems near Roche-lobe filling. On the other hand, the off-resonant approximation shows that the dynamical tide can contribute to $\sim$20\% of the precession for the orbits with small eccentricities at close separations. We also study the effect of the WD rotations on the precession rate. The Coriolis force from the rotation of the WDs has a suppression effect on the dynamical tide precession and can also cause negative precession if the spin is high enough. Meanwhile, the centrifugal force induces a quadrupolar deformation of the WDs which gives extra positive precession comparable to the equilibrium tide effect when the system is at a pseudo-synchronous state, as shown in Willems \textit{et al.} \cite{PhysRevLett.100.041102}.

We also study the effect of the dynamical tide precession on the GW signal. Assuming a 4-year signal duration, we show the parameter space where the dynamical tide precession is resolvable by LISA. 
Compared to the frequency shift due to 2.5PN radiation reaction in a (0.2, 0.6)~$M_\odot$ system, this effect is resolvable within a larger range of orbital parameters. At close separations, it causes $\gtrsim10$\% systematic shift in the eccentricity measurement in less eccentric systems since the precession rate depends on both the eccentricity and the tidal parameters.
For the case with (0.6, 0.6)~$M_\odot$, the resolvable parameter range of the 2.5PN effect is larger than that of the effect of the dynamical tide except at resonance. Hence, a stronger chirp effect is expected in systems with higher masses.
The highly eccentric systems at close separations can lie within the chaotic regime, where the dynamical tide causes the orbit to evolve chaotically, a phenomenon first illustrated by Mardling \cite{1995ApJ...450..732M}. This produces an unpredictable GW signal that poses problems in the detection.

Assuming a non-chaotic signal, we show that the dynamical tidal effect in the precession rate can be distinguished from other factors by analysing the waveform if the chirp from the 2.5PN effect is strong enough, or if the system is at resonance, given a high enough SNR. Therefore, we conclude that LISA can measure the dynamical tide within high mass eccentric DWD systems or the low mass systems at resonance.

Regarding future work, a more detailed waveform analysis of the influence of the precession effect on the signal is required to analyze the measurement error in parameter estimation. Besides, the effect of the $g$-modes on the orbital motion and evolution of the eccentric DWDs can also be an interesting direction to pursue. We anticipate that the multiple harmonics in eccentric systems would lead to a richer resonance behaviour than in the circular case. This allows us to potentially probe the dissipation mechanisms and learn more about the interior structure of the WDs from the excitation and damping of the modes.

\acknowledgments
We all acknowledge support from NASA Grant No. 80NSSC20K0523. 
K.Y. acknowledges support from a Sloan Foundation Research Fellowship and the Owens Family Foundation. 
K.Y. would like to also acknowledge support by the COST Action GWverse CA16104 and JSPS KAKENHI Grant No. JP17H06358.
\appendix

\section{\label{Sec:waveform} Gravitational waveform models}

\subsection{\label{SSec:Numerical waveform}The numerical waveform}
The far-field metric perturbation from the leading order GW emission is given by the quadrupole formula
\begin{align}
    h_{ij} = \frac{G}{c^4}\frac{2 \ddot{Q}_{ij}}{d}, \label{eq:quad_formula}
\end{align}
where $c$ is the speed of light, $Q_{ij}$ is the quadrupole moment of the source while $d$ is the distance between the source and the point of observation. We consider only the quadrupole moment of the orbit and ignore the effects from the non-radial deformations of the individual WDs.

The orbit is governed by the equation
\begin{align}
    \mathbf{a} = -\frac{GM}{D^2}\mathbf{n} + \mathbf{a}_\text{1PN} + \mathbf{a}_\text{tide}, \label{eq:eom}
\end{align}
where $\mathbf{a}_\text{1PN}$ is the acceleration due to the 1PN effect while the tidal acceleration term $\mathbf{a}_\text{tide}$ is given in Eq.~\eqref{eq:tide_acc}. The dissipative effects like radiation reaction are ignored here.
The 1PN effect is given by the Einstein-Infeld-Hoffmann equation \cite{10.2307/1968714}:
\begin{align}
    \mathbf{a}_\text{1PN} =& - \frac{GM}{D^2 c^2} \Bigg\{ \bigg[ \left(1+3\eta\right)v^2-\frac{3}{2}\eta \left(\mathbf{n}\cdot \mathbf{v}\right)^2 \nonumber \\   &-2\left(2+\eta\right)\frac{GM}{D}\bigg]\mathbf{n}
    -2\left(2-\eta\right)\left(\mathbf{n}\cdot \mathbf{v}\right)\mathbf{v}\Bigg\},
\end{align}
where $\eta=m_1 m_2 /(m_1 m_2)^2$ is the symmetric mass ratio and $\mathbf{v}$ is the relative velocity.

Equation~\eqref{eq:eom} is numerically integrated and the result is substituted into Eq.~\eqref{eq:quad_formula} to obtain the quadrupolar waveform of the plus and cross polarizations, $h_+$ and $h_\times$ in the transverse-traceless gauge. The strain signal detected is written as
\begin{align}
    h_{j}(t) =& F_{+}^{j}(t) h_+(t) + F_{\times}^{j}(t) h_\times(t), \label{eq:h_I}
\end{align}
where $j=I,II$, and $F_{+}^{j}$, $F_{\times}^{j}$ are the antenna pattern functions (see \cite{PhysRevLett.87.251101,10.1093/mnras/289.1.185,PhysRevD.69.082005,PhysRevD.57.7089}). The functions $F_{+}^{j}$ and $F_{\times}^{j}$ depend on the source's angular position ($\theta_S$, $\phi_S$) and its orientation ($\theta_L$, $\phi_L$) in the ecliptic coordinate system. For simplicity, we set $\theta_L = \pi/4$ and the rest as zero in this study.

\subsection{\label{SSec:PM_waveform} Peters and Mathews waveform}

The Peters and Mathews waveform \cite{PhysRev.131.435, PhysRevD.69.082005} describes the leading order gravitational radiation emitted from a binary system in an eccentric Keplerian orbit. In this section, we write down the formulation of the waveform used in \cite{PhysRev.131.435}, without including the full evolution equations of the orbital elements.

Choosing the orbital plane as the $x$-$y$ plane of the coordinate system, with the pericenter lying on the positive $x$-axis initially, the plus and cross polarizations  are written as a sum of harmonics
\begin{align}
    h_+ =& \frac{G}{c^4 d} \sum_k \bigg\{\left(1+\cos^2\iota\right)\Big[C_a^{(k)} \cos{(k\Phi)} \cos(2\gamma) \nonumber \\
    &- C_b^{(k)} \sin{(k\Phi)} \sin(2\gamma)\Big]+ C_c^{(k)} \cos{(k\Phi)} \sin^2\iota\bigg\}, \label{eq:hp}\\
    h_\times =& \frac{2 \cos\iota}{d} \frac{G}{c^4} \sum_k \Big[C_a^{(k)} \cos{(k\Phi)} \sin(2\gamma) \nonumber \\
    &+ C_b^{(k)} \sin{(k\Phi)} \cos(2\gamma)\Big] \label{eq:hc},
\end{align}
where $\iota$ is the inclination angle of the source, given by
\begin{align}
    \cos \iota = \cos\theta_L \cos\theta_S +\sin\theta_L \sin\theta_S \cos{(\phi_L-\phi_S)},
\end{align}
and the precession is given by $\gamma =\dot{\gamma} t$ and $C_i^{(k)}$, with $i=a,b,c$, are the Fourier coefficients of the quadrupole moment components:
\begin{align}
\frac{1}{2}\left(\ddot{Q}_{11}-\ddot{Q}_{22}\right) =& \sum_{k=0}^\infty C_a^{(k)} \cos{k \Phi(t)},\\
\ddot{Q}_{12} =& \sum_{k=0}^\infty C_b^{(k)} \sin{k \Phi(t)},\\
\frac{1}{2}\left(\ddot{Q}_{11}+\ddot{Q}_{22}\right) =& \sum_{k=0}^\infty C_c^{(k)} \cos{k \Phi(t)}.
\end{align}
In the Keplerian orbits, the coefficients can be written in terms of the Bessel functions \cite{PhysRevD.69.082005}:
\begin{align}
    C_a^{(k)} =& -k\mu \left(\frac{G\Omega M}{c^3}\right)^{2/3} \Big[J_{k-2}(k e) - 2eJ_{k-1}(k e) \nonumber \\
    & + \frac{2}{k}J_{k}(k e)+ 2e J_{k+1}(k e) - J_{k+2}(k e)\Big],\\
    C_b^{(k)} =& -k\mu \left(\frac{G\Omega M}{c^3}\right)^{2/3} \sqrt{1 - e^2} \Big[J_{k-2}(k e) & \nonumber\\
    &- 2J_{k}(k e) + J_{k+2}(k e)\Big], \\
    C_c^{(k)} =& 2\mu \left(\frac{G\Omega M}{c^3}\right)^{2/3}J_{k}(k e),
\end{align}
where $\mu$ is the reduced mass of the binary. The effect of the tide and 1PN correction are included only through $\dot{\gamma}$. The amplitude corrections are not included as they are expected to be small within the parameter range of interest. The waveform model serves as a good approximation as long as $\dot{\gamma}$ is much smaller than $\Omega$ \cite{10.1093/mnras/274.1.115}.

\begin{figure}
    \includegraphics[width=8.6cm]{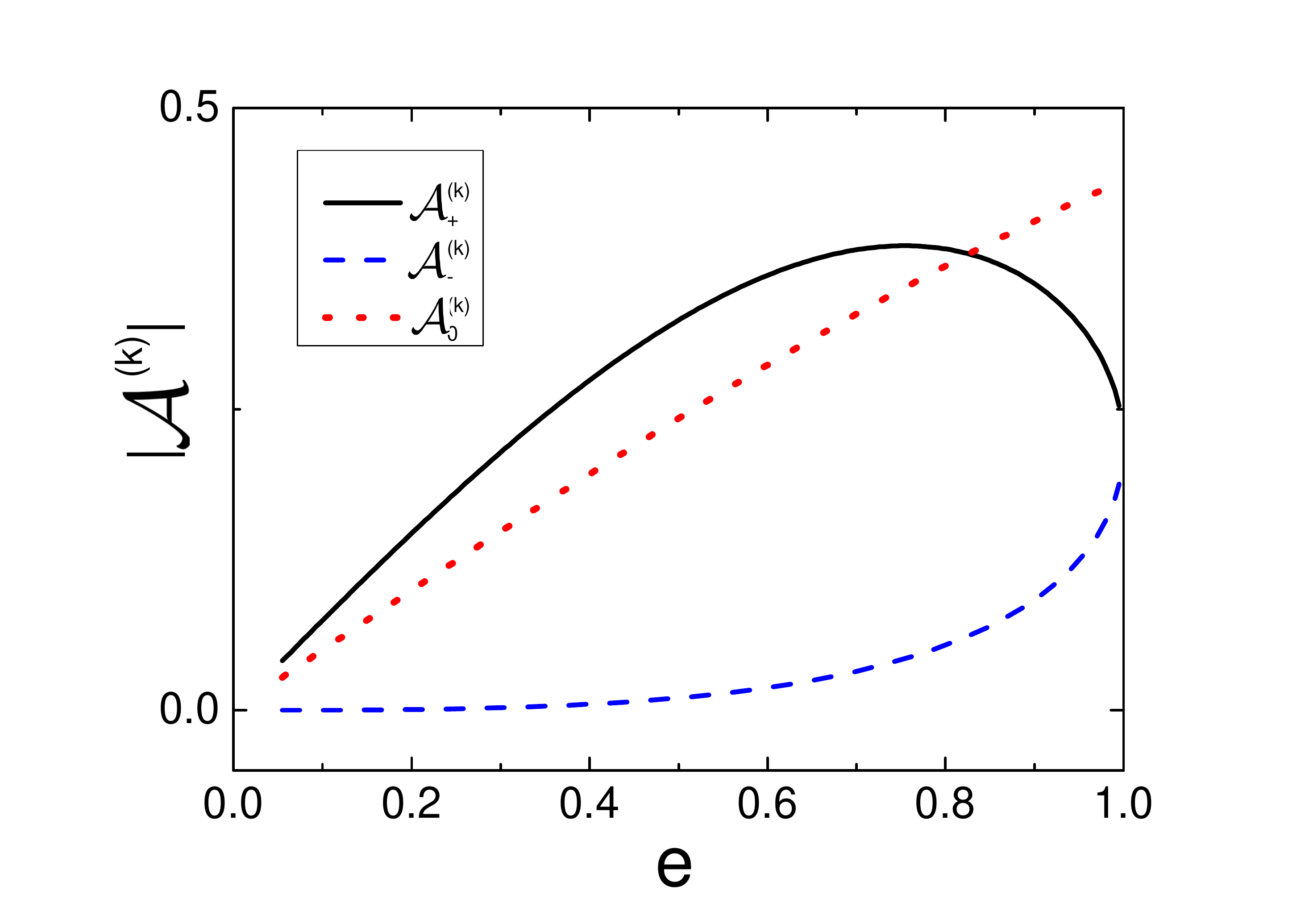}
    \caption{\label{fig:ecc_wf_amp} The waveform amplitudes  $\mathcal{A}^{(k)}_+$, $\mathcal{A}^{(k)}_-$ and $\mathcal{A}^{(k)}_0$ at different eccentricities for the $k=1$ harmonic.}
\end{figure}

Without chirping, the precession effect splits each harmonic of the waveform into three distinct frequencies at $k\Omega$ and $k\Omega\pm 2\dot{\gamma}$. This can be illustrated by re-expressing Eqs.~\eqref{eq:hp} and \eqref{eq:hc} in terms of $\mathcal{A}^{(k)}_\pm \cos{(k \Omega t\pm2\gamma)}$ and $\mathcal{A}^{(k)}_0 \cos{(k\Omega t)}$ as in \cite{10.1093/mnras/274.1.115,PhysRevLett.87.251101}, where the amplitudes $\mathcal{A}^{(k)}_\pm$ and $\mathcal{A}^{(k)}_0 $ are $(C_a^{(k)}\pm C_b^{(k)})/2$ and $C_c^{(k)}$ without the prefactor $\mu \left(G\Omega M/c^3\right)^{2/3}$. Using the series expansion of the Bessel functions, these amplitude terms can be shown to scale as $\mathcal{A}^{(k)}_+ \sim e^{k-2}$ for $k\ne 1$ and $\mathcal{A}^{(k)}_+ \sim e$ for $k=1$, $\mathcal{A}^{(k)}_- \sim e^{k+2}$, and $\mathcal{A}^{(k)}_0 \sim e^{k}$. Therefore, $\mathcal{A}^{(k)}_+$ has the largest amplitude while $\mathcal{A}^{(k)}_-$ is the smallest out of the three harmonics for low to intermediate eccentricities. The amplitude functions for the $k=1$ harmonic at different eccentricities are shown in Fig.~\ref{fig:ecc_wf_amp}. Except for large eccentricities ($e \gtrsim 0.5$), the relation $|\mathcal{A}^{(k)}_+| >|\mathcal{A}^{(k)}_0| >|\mathcal{A}^{(k)}_-|$ holds for all harmonics \cite{PhysRevLett.100.041102}.

The chirping can be included through expanding the phase of each harmonic  as $\Phi(t) = \Omega t + \dot{\Omega}t^2/2$, where $\dot{\Omega}$ contains the 2.5PN radiation reaction effect \cite{PhysRev.131.435} given by
\begin{align}
    \dot{\Omega} =& \frac{96}{5} \frac{c^6}{G^2}\frac{\mu}{M^3} (1-e^2)^{-7/2} \left(\frac{GM\Omega}{c^3}\right)^{11/3} \nonumber \\
    &\times \Big(1+\frac{73}{24}e^2 + \frac{37}{96}e^4\Big).
\end{align}
The Doppler phase term due to LISA's motion (see \cite{PhysRevD.57.7089}) is not included in $\Phi(t)$ since we set $\theta_S = 0$.

\nocite{*}

\bibliography{apssamp}

\end{document}